%% file: POR4ATL-Paradoxes-KR2021-final-arxiv.tex
\PassOptionsToPackage{dvipsnames,svgnames,table}{xcolor}

\documentclass{article}
\usepackage{kr}

\usepackage{times}
\usepackage{soul}
\usepackage{url}
\usepackage[hidelinks]{hyperref}
\usepackage[utf8]{inputenc}
\usepackage[small]{caption}
\usepackage{graphicx}
\usepackage{amsmath}
\usepackage{amsthm}
\usepackage{booktabs}
\usepackage{balance} 
\usepackage{xspace}
\usepackage[super]{nth}  
\usepackage{tikz}
\usepackage{subcaption}
\usepackage{eurosym}
\usepackage[normalem]{ulem}
\usepackage{diagbox}
\usepackage{pifont}
\usepackage{mathtools}
\usepackage{etoolbox}
\usepackage[capitalise,english,nameinlink]{cleveref}
\usepackage{soul}
\usepackage{url}
\usepackage[utf8]{inputenc}
\usepackage{amsfonts,amssymb}
\usepackage{graphicx}
\usepackage{booktabs}
\usepackage{epsf}
\usepackage{stmaryrd}
\usepackage{longtable}
\usepackage{latexsym}
\usepackage{color}
\usepackage[mathscr]{eucal}
\usepackage[refpage,english]{nomencl}
\usepackage{xspace}
\usepackage{wrapfig}
\usepackage{url}
\usetikzlibrary{arrows,automata,calc,shapes,decorations,backgrounds,petri,mindmap,fit,positioning}
\usepackage{wojtek15logics,wojtek15logicsWP,wojtek15other}
\usepackage{macros}

\newtheorem{thm}{Theorem}[section]
\numberwithin{thm}{section}

\newtheorem{smalltheorem}[thm]{Proposition}
\newtheorem{example}[thm]{Example}
\newtheorem{theorem}[thm]{Theorem}
\newtheorem{lemma}[thm]{Lemma}

\newtheorem{definition}[thm]{Definition}
\definecolor[named]{tucgreen}{RGB}{0,140,79}

\newcommand{\extended}[1]{}    
\newcommand{\short}[1]{#1}     
\renewcommand{\extended}[1]{#1} 
\renewcommand{\short}[1]{}      

\newcommand{\eps}{{\text{\normalsize$\epsilon$}}}
\newcommand{\stratevt}[1]{\pmb{#1}}

\newcommand{\IISEps}{IIS^\eps}

\newcommand{\AMAS}{S\xspace}
\newcommand{\AMASVoting}{S_{vote}\xspace}
\newcommand{\epsAMAS}{\ensuremath{S^\eps}\xspace}
\newcommand{\ASV}[2]{\ensuremath{\mathrm{ASV}_{#1,#2}}}
\newcommand{\fullmodel}{\mathit{IIS}}
\newcommand{\CF}{\textbf{CF}\xspace}
\newcommand{\SCF}{\textbf{SCF}\xspace}
\newcommand{\EL}{\textup{React}\xspace}
\renewcommand{\AE}{\textbf{AE}\xspace}

\renewcommand{\PV}{\mathcal{PV}}
\newcommand{\hatPV}{\mathit{PV}} 
\newcommand{\state}{g}
\renewcommand{\model}{M}
\newcommand{\modelEpsilon}{\model^\eps}

\newcommand{\modelEps}{\model}
\newcommand{\modelEpsPrim}{{\model{'}}}
\newcommand{\outcomeEps}{\outcome^{\EL}}
\newcommand{\outcomeCF}{\outcome^{\CF}}
\newcommand{\outcomeSCF}{\outcome^{\SCF}}
\newcommand{\outcomeEpsCF}{\outcome^{\EL,\CF}}

\renewcommand{\A}{\Agt}

\newcommand{\events}{\mathit{Evt}}  
\newcommand{\evt}{\alpha}            
\newcommand{\evttwo}{\beta}            
\newcommand{\private}{black}
\newcommand{\Mconf}{M_\mathit{conf}}

\newcommand{\roc}{R}
\newcommand{\satisfair}[2][]{\satisf[#1]^{#2}}

\title{Strategic Abilities of Asynchronous Agents: \\ Semantic Side Effects and How to Tame Them}

\author{
Wojciech Jamroga$^{1,2}$\and
Wojciech Penczek$^1$\and
Teofil Sidoruk$^{1,3}$ \\
\affiliations
$^1$Institute of Computer Science, Polish Academy of Sciences, Warsaw, Poland\\
$^2$Interdisciplinary Centre on Security, Reliability and Trust, SnT, University of Luxembourg\\
$^3$Faculty of Mathematics and Information Science, Warsaw University of Technology\\
\emails
\{jamroga, penczek, t.sidoruk\}@ipipan.waw.pl
}

\begin{document}

\maketitle

\begin{abstract}
Recently, we have proposed a framework for verification of agents' abilities in asynchronous multi-agent systems (MAS), together with an algorithm for automated reduction of models~\cite{Jamroga18por}.
The semantics was built on the modeling tradition of distributed systems.
As we show here, this can sometimes lead to counterintuitive interpretation of formulas when reasoning about the outcome of strategies.
First, the semantics disregards finite paths, and yields unnatural evaluation of strategies with deadlocks.
Secondly, the semantic representations do not allow to capture the asymmetry between proactive agents and the recipients of their choices.
We propose how to avoid the problems by a suitable extension of the representations and change of the execution semantics for asynchronous MAS.
We also prove that the model reduction scheme still works in the modified framework.
\end{abstract}


\section{Introduction}\label{sec:intro}
\input{Introduction.tex}

\section{Models of Multi-agent Systems}\label{sec:prelim}
\input{Preliminaries.tex}

\section{Reasoning About Abilities: ATL*}\label{sec:atl}
\input{Reasoning.tex}

\section{\mbox{Semantic Problems and How to Avoid Them}}\label{sec:paradoxes}
\input{Paradoxes}

\section{Partial Order Reduction Still Works}\label{sec:por}
\input{Po-reduction-short.tex}

\section{Conclusions}\label{sec:conclusions}
\input{Conclusions.tex}


\section*{Acknowledgements}
We thank the anonymous reviewers for their insightful comments.
The authors acknowledge the support of the National Centre for Research and Development, Poland (NCBR),
and the Luxembourg National Research Fund (FNR), under the PolLux/FNR-CORE project STV (POLLUX-VII/1/2019).
W. Penczek and T. Sidoruk acknowledge support from CNRS/PAS project PARTIES.


\balance
\bibliographystyle{kr}
\bibliography{bib,wojtek,wojtek-own}


\clearpage
\appendix

\section{Partial Order Reduction: Details}\label{sec:por-details}
\input{Po-reduction-v3.tex}

\end{document}

%% file: Introduction.tex
\short{\nocite{Alur97ATL,Alur01jmocha,Kacprzak04umc-atl,Lomuscio06mcmas,Pilecki14synthesis,peled96b,gw,GKPP99,LomuscioPQ10a,PenczekSGK00}}
\extended{\para{Modal logics of strategic ability.}}
\emph{Alter\-nating-time temporal logic} \ATLs~\short{\cite{Alur02ATL,Schobbens04ATL}}\extended{\cite{Alur97ATL,Alur02ATL,Schobbens04ATL}} is probably the most popular logic
to describe interaction\extended{ of agents} in multi-agent systems.
Formulas of \ATLs allow to express statements about what agents (or groups of agents) can achieve.
For example, $\coop{taxi}\Always\neg\prop{fatality}$ says that the autonomous cab can drive in such a way that nobody is ever killed, and $\coop{taxi,passg}\Sometm\prop{destination}$ expresses that the cab and the passenger have a joint strategy to arrive at the destination, no matter what any other agents do.
Such statements allow to express important functionality and safety requirements in a simple and intuitive way.
Moreover, the provide input to algorithms and tools for verification\extended{ of strategic abilities}, that have been in constant development for over 20 years~
\short{\cite{Alur98mocha-cav,Chen13prismgames,Busard14improving,Huang14symbolic-epist,Cermak14mcheckSL,Lomuscio15mcmas,CermakLM15,BelardinelliLMR17a,BelardinelliLMR17b,Jamroga19fixpApprox-aij,Kurpiewski21stv-demo}}%
\extended{\cite{Alur98mocha-cav,Alur01jmocha,Kacprzak04umc-atl,Lomuscio06mcmas,Chen13prismgames,Busard14improving,Pilecki14synthesis,Huang14symbolic-epist,Cermak14mcheckSL,Lomuscio15mcmas,CermakLM15,BelardinelliLMR17a,BelardinelliLMR17b,Jamroga19fixpApprox-aij,Kurpiewski19stv-demo,Kurpiewski21stv-demo}}.
\extended{
  Still, there are two caveats.

  First, all the realistic scenarios of agent interaction, that one may want to specify and verify, involve imperfect information.
  That is, the agents in the system do not always know exactly the global state of the system, and thus they have to make their decisions based on their local view of the situation.
  Unfortunately, verification of agents with imperfect information is hard to very hard -- more precisely, \Deltwo-complete to undecidable, depending on the syntactic and semantic variant of the logic~\cite{Schobbens04ATL,Guelev11atl-distrknowldge,Dima11undecidable}.
  Also, the imperfect information semantics of \ATLs does not admit alternation-free fixpoint
  characterizations~\cite{Bulling11mu-ijcai,Dima14mucalc,Dima15fallmu}, which makes incremental synthesis
  of strategies\extended{ impossible, or at least} difficult to achieve~\cite{Pilecki14synthesis,Busard14improving,Huang14symbolic-epist,Busard15reasoning,Jamroga17fixpApprox}.
}

\short{\para{Asynchronous semantics and partial-order reduction.}}
\extended{Secondly, t}\short{T}he semantics of strategic logics is traditionally based on synchronous concurrent game models.
\extended{
  In other words, one implicitly assumes the existence of a global clock that triggers subsequent global events in the system;
  at each tick of the clock, all the agents choose their actions, and the system proceeds accordingly with a global transition.
}%
However, many real-life systems are inherently asynchronous\extended{, and do not operate on a global clock that perfectly synchronizes the atomic steps of all the components. Moreover, many systems that are synchronous at the implementation level}\short{ or} can be more conveniently modeled as asynchronous\extended{ on a more abstract level}.
\extended{
  In many scenarios, both aspects combine.
  For example, when modeling an anti-poaching operation~\cite{Fang17anti-poaching}, one may take into account the truly asynchronous nature of events happening in different national parks, but also the best level of granularity for modeling the events happening within a single nature reserve.}

\extended{\para{Asynchronous semantics and partial-order reduction.}}
We have recently proposed how to adapt the semantics of \ATLs to asynchronous MAS~\cite{Jamroga18por}.
We also showed that the technique of \emph{partial order reduction (POR)}~\short{\cite{peled-representatives,peled-on_the_fly,LomuscioPQ10b}}%
\extended{\cite{peled-representatives,peled-on_the_fly,peled96b,gw,GKPP99,LomuscioPQ10a,LomuscioPQ10b}} can be adapted to verification of strategic abilities in asynchronous MAS.
In fact, the (almost 30 years old) POR for linear time logic \LTL can be taken off the shelf and applied to a significant part of \ATLsir,
the variant of \ATLs based on strategies with imperfect information and imperfect recall.
This is very important, as the practical verification of asynchronous systems is often impossible due to the state- and transition-space explosion resulting from interleaving of local transitions.
POR allows for a significant, sometimes even exponential, reduction of the models.

\para{Semantic side effects.}
While the result is appealing, there is a sting in its tail: the \ATLs semantics in~\cite{Jamroga18por} leads to counterintuitive interpretation of strategic properties.
First, it disregards finite paths, and evaluates some intuitively losing strategies as winning (and vice versa).
Secondly, it provides a flawed interpretation of the \emph{concurrency fairness} assumption.
Thirdly, the representations and their execution semantics do not allow to capture the asymmetry between the agents that control which synchronization branch will be taken, and those influenced by their choices.
We tentatively indicated some of the problems in the extended abstract~\cite{Jamroga21paradoxes-ea}. In this paper, we demonstrate them carefully, and propose how they can be avoided.

\para{Contribution.}
Our contribution is threefold.
First, we discuss in detail the semantic side effects of adding strategic reasoning on top of classical models of concurrent systems~\cite{Priese83apa-nets}.
We identify the reasons, and demonstrate the problematic phenomena on simple examples.
Secondly, we show how to avoid these pitfalls by extending the class of representations and slightly changing the execution semantics of strategies.
Specifically, we add ``silent'' $\epsilon$-transitions in the models and on outcome paths of strategies, and allow for nondeterministic choices in the agents' repertoires.
We also identify a family of fairness-style conditions, suitable for the interaction of proactive and reactive agents.
No less importantly, we prove that partial order reduction is still correct in the modified framework.

\para{Motivation.}
The variant of \ATLs for asynchronous systems in~\cite{Jamroga18por} was proposed mainly as a framework for formal verification. This was backed by the results showing that it submits to partial order reduction. 
However, a verification framework is only useful if it allows to specify requirements in an intuitive way, so that the property we \emph{think} we are verifying is indeed \emph{the one being verified}. In this paper, we show that this was not the case. We also propose how to overcome the problems without spoiling the efficient reduction scheme. The solutions are not merely technical. In fact, they lead to a better understanding of how strategic activity influences the overall behavior of the system, and how it should be integrated with the traditional models of asynchronous interaction.

%% file: Preliminaries.tex

We first recall the models of asynchronous interaction in MAS, proposed in~\cite{Jamroga18por}
and inspired by~\short{\cite{Priese83apa-nets,LomuscioPQ10b}}\extended{\cite{Priese83apa-nets,Fagin95knowledge,LomuscioPQ10b}}.

\short{\para{Asynchronous multi-agent systems.}}\extended{\subsection{Asynchronous Multi-agent Systems}}\label{sec:asynchr}
In logical approaches to MAS, one usually assumes synchronous actions of all the agents~\cite{Alur02ATL,Schobbens04ATL}.
However, many agent systems are inherently asynchronous, or it is useful to model them without assuming precise
timing relationships between the actions of different agents.
\extended{
  As an example, consider a team of logistic robots running in a factory~\cite{Schlingloff16embedded}.
  Often no global clock is available to all the robots, and even if there is one, the precise relative timing
  for robots operating in different places is usually irrelevant.

}
Such a system can be conveniently represented with a set of automata that execute asynchronously by interleaving
local transitions, and synchronize their moves whenever a shared event occurs.
The idea is to represent the behavior of each agent by a finite automaton where the nodes and transitions correspond,
respectively, to the agent's local states and the events in which it can take part.
Then, the global behavior of the system is obtained by the interleaving of local transitions,
assuming that, in order for a shared event to occur, all the corresponding agents must execute it in their automata.
%
This motivates the following definition.

\begin{definition}[Asynchronous MAS]\label{def:amas}
An \emph{asynchronous multi-agent system (AMAS)} \AMAS consists of $n$ agents $\A = \set{1,\dots,n}$,\extended{\footnote{
  We do not consider the environment component, which may be added with no technical difficulty.}}
each associated with a tuple $A_i =(L_i, \iota_i, \events_i, \roc_i, T_i{,\PV_i,V_i})$ including a set
of \emph{\extended{possible }local states} $L_i=\{l_i^1, l_i^2,\dots,l_i^{n_i}\}$,
an \emph{initial state} $\iota_i\in L_i$, and
a set of \emph{events} $\events_i=\{\evt_i^1,\evt_i^2,\ldots, \evt_i^{m_i}\}$.
An agent's \emph{repertoire of choices}\extended{\footnote{
  In interpreted systems, this function is usually referred to as a \emph{protocol}. Here, we opt for a different name to avoid possible confusion, e.g., with security protocols.}}
$\roc_i: L_i \to 2^{\events_i}\setminus\{\emptyset\}$ selects the events available at each local state.
$T_i: L_i \times \events_i \fpart L_i$ is a (partial) \emph{local transition function} such that
$T_i(l_i,\evt)$ is defined iff $\evt\in \roc_i(l_i)$.
That is, $T_i(l,\evt)$ indicates the result of executing event $\evt$ in local state $l$ from the perspective of agent $i$.

Let $\events = \bigcup_{i \in \A} \events_i$ be the set of all events, and
$Loc = \bigcup_{i \in \A} L_i$ be the set of all local states in the system.
For each event $\evt \in \events$, $Agent(\evt) = \set{i\in\A \mid \evt \in \events_i}$ is the set of agents which have $\evt$ in their repertoires;
events shared by multiple agents are jointly executed by all of them.
We assume that each agent $i$ in the AMAS is endowed with a disjoint set of its \emph{local propositions $\PV_i$}, and their valuation $V_i : L_i \then \powerset{\PV_i}$.
The overall set of propositions $\PV = \bigcup_{i\in\A} \PV_i$ collects all the local propositions.
\end{definition}

As our working example, we use the following scenario.

\begin{example}[Conference in times of epidemic]\label{ex:conference-amas}
Consider the AMAS in Figure~\ref{fig:conference}, consisting of the Steering Committee Chair ($sc$), the General Chair ($gc$), and the Organizing Committee Chair ($oc$).
Faced with the Covid-19 epidemics, $sc$ can decide to give up the conference, or send a signal to $gc$ to proceed and open the meeting.
Then, $gc$ and $oc$ jointly decide whether the conference will be run on site or online.
In the former case, the epidemiologic risk is obviously much higher, indicated by the atomic proposition \prop{epid}.

The set of events, the agents' repertoires of choices, and the valuation of atomic propositions can be easily read from the graph.
\extended{For easier reading, all the private events are shown in grey.}
Note that event $proceed$ is shared by agents $sc$ and $gc$, and can only be executed jointly.
Similarly, $onsite$ and $online$ are shared by $gc$ and $oc$.
All the other events are private, and do not require synchronization.
\end{example}

\newcommand{\scale}{0.75}
\begin{figure}[t]
\centering
\begin{tabular}{@{}c@{\quad}c@{\;}c@{}}
 \textbf{gc} & \textbf{oc} & \textbf{sc} \\ \\
	\begin{tabular}{@{}c@{}}
	\begin{tikzpicture}[->,>=stealth',shorten >=1pt,auto,node distance=2.1cm,transform shape,semithick,scale=\scale]
     \input{conference-gc+strat.tex}
    \end{tikzpicture}
    \end{tabular}
 &
  \begin{tabular}{@{}c@{}}
	\begin{tikzpicture}[->,>=stealth',shorten >=1pt,auto,node distance=2.1cm,transform shape,semithick,scale=\scale]
     \input{conference-oc-2+strat.tex}
  \end{tikzpicture}
  \end{tabular}
 &
	\begin{tabular}{@{}c@{}}
    \begin{tikzpicture}[->,>=stealth',shorten >=1pt,auto,node distance=2.1cm,transform shape,semithick,scale=\scale]
     \input{conference-sc+strat.tex}
    \end{tikzpicture}
    \end{tabular}
\end{tabular}
\caption{Simple asynchronous MAS: agents $gc$, $oc$, and $sc$. A joint strategy of agents $\set{gc,oc}$ is highlighted. }
\label{fig:conference}\label{fig:conference-strategy}
\end{figure}
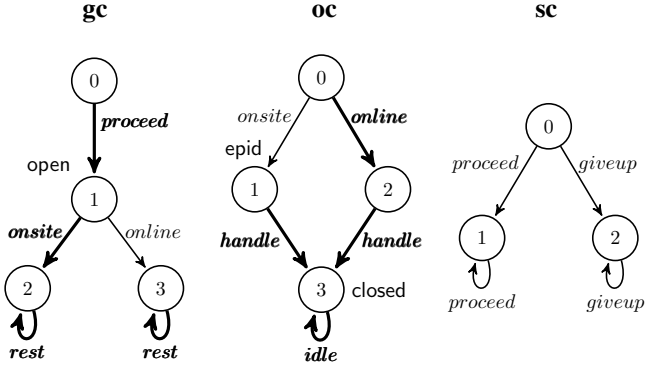
\renewcommand{\scale}{0.85}

\short{\para{Interleaved interpreted systems.}}\extended{\subsection{Interleaved Interpreted Systems}}
\label{sec:iis}
To understand the interaction between asynchronous agents, we use the standard execution semantics from concurrency models, i.e., interleaving with synchronization on shared events. To this end, we compose the network of local automata (i.e., AMAS) to a single automaton based on the notions of \emph{global states} and \emph{global transitions}, see below.

{
\begin{definition}[Model]\label{def:canonical-iis}
Let $\AMAS$ be an AMAS with $n$ agents.
Its \emph{model} $IIS(\AMAS)$ extends $\AMAS$ with:
(i) the set of global states $\States \subseteq L_1\times\ldots\times L_n$, including the \emph{initial state} $\iota = (\iota_1,\dots,\iota_n)$ and all the states reachable from $\iota$ by $T$ (see below);\
(ii) the \emph{global transition function} $T: \States\times \events \fpart \States$, defined by $T(\state_1,\evt)= \state_2$ iff $T_i(\state_1^i,\evt) = \state^i_2$
for all $i \in Agent(\evt)$ and $\state_1^i = \state^i_2$ for all $i \in \A \setminus Agent(\evt)$;\
(iii) the \emph{global valuation} of propositions $V: \States \rightarrow 2^{\PV}$, defined as $V(l_1,\dots,l_n) = \bigcup_{i\in\A} V_i(l_i)$.
\end{definition}
}

Models, sometimes called \emph{interleaved interpreted systems} (IIS), are used to provide an execution semantics to AMAS\extended{, and consequently provide us with semantic structures to reason about AMAS}.
Intuitively, the global states in $IIS(\AMAS)$ can be seen as the possible configurations of local states of all the agents.
Moreover, the transitions are labeled by events that are simultaneously selected (in the current configuration) by all the agents that have the event in their repertoire.
\extended{Clearly, private events (i.e., events such that $Agent(\evt)$ is a singleton) require no synchronization.}

\begin{example}[Conference]\label{ex:conference-iis}
The model for the asynchronous MAS of Example~\ref{ex:conference-amas} is shown in Figure~\short{\ref{fig:conference-model}}\extended{\ref{fig:conference}}.
\end{example}

We say that event $\evt \in \events$ is \emph{enabled} at $\state\in \States$ if $T(\state,\evt)= \state'$ 
for some $\state' \in \States$.
The set of events enabled at $\state$ is denoted by $enabled(\state)$.
The global transition function is assumed to be serial, i.e., at each $\state\in \States$ there
exists at least one enabled event.

\para{Discussion.}
This modeling approach is standard in {theory of concurrent systems}, where it dates back to the early 1980s and the idea of APA Nets (asynchronous, parallel automata nets)~\cite{Priese83apa-nets}.
Note that APA Nets and their models were \emph{not} proposed with causal interpretation in mind.
In particular, they were \emph{not} meant to capture the interaction of purposeful agents that freely choose their strategies, but rather a set of reactive components converging to a joint behavior.
Despite superficial differences, the same applies to process-algebraic approaches to concurrency, such as CSP~\cite{Hoare78csp}, CCS~\cite{Milner80ccs}, ACP~\cite{Bergstra85acp}, and $\pi$-calculus~\cite{Milner92picalculus}.

Definition~\ref{def:amas} extends that with the repertoire functions from synchronous models of MAS~\cite{Lomuscio00knowledge,Alur02ATL}.
Agent $i$'s repertoire lists the events available to $i$, and is supposed to define the space of $i$'s strategies. As we show further, this is not enough in case of asynchronous MAS.

%% file: conference-gc+strat.tex
\tikzstyle{every state}=[fill=none,draw=black,text=black,minimum size=0.8cm]
\tikzstyle{privlabel}=[]
\tikzstyle{privtrans}=[]
\tikzstyle{strat}=[very thick]

\node[state] (s0) {$0$}; 
\node[state] (s1) [below of=s0, label=above left:{\large $\prop{open}$}] {$1$};
\node[state] (s2) [below left=1cm and 0.6cm of s1] {$2$};
\node[state] (s3) [below right=1cm and 0.6cm of s1] {$3$};

\path
(s0)
  edge[strat] node[near start,right] {$\stratevt{proceed}$}	(s1)
(s1)
  edge[strat] node[near start,left] {$\stratevt{onsite}$} (s2)
  edge node[near start,right] {${online}$} (s3)
(s2)
  edge[loop below,strat,privtrans] node[privlabel,midway,below] {$\stratevt{rest}$} (s2)
(s3)
  edge [loop below,strat,privtrans] node[privlabel,midway,below] {$\stratevt{rest}$} (s3);

%% file: conference-oc-2+strat.tex
\tikzstyle{every state}=[fill=none,draw=black,text=black,minimum size=0.8cm]
\tikzstyle{privlabel}=[]
\tikzstyle{privtrans}=[]
\tikzstyle{strat}=[very thick]

\node[state] (s0) {$0$}; 
\node[state] (s1) [below left=1.4cm and 0.6cm of s0, label=above:{\large $\prop{epid}\quad{}$}] {$1$};
\node[state] (s2) [below right=1.4cm and 0.6cm of s0] {$2$};
\node[state] (s3) [below right=1.2cm and 0.6cm of s1, label=right:{\large $\prop{closed}\quad{}$}] {$3$};

\path
(s0)
  edge node[near start,left] {$onsite$} (s1)
  edge[strat] node[near start,right] {$\stratevt{online}$}	(s2)
(s1)
  edge [strat,privtrans] node[privlabel,midway,left] {$\stratevt{handle}$} (s3)
(s2)
  edge [strat,privtrans] node[privlabel,midway,right] {$\stratevt{handle}$} (s3)
(s3)
  edge [strat,privtrans,loop below] node[privlabel,midway,below] {$\stratevt{idle}$} (s3);

%% file: conference-sc+strat.tex
\tikzstyle{every state}=[fill=none,draw=black,text=black,minimum size=0.8cm]
\tikzstyle{privlabel}=[]
\tikzstyle{privtrans}=[]

\node[state] (s0) {$0$}; 
\node[state] (s1) [below left=1.4cm and 0.6cm of s0] {$1$};
\node[state] (s2) [below right=1.4cm and 0.6cm of s0] {$2$};

\path
(s0)
  edge node[near start,left] {${proceed}$} (s1)
  edge[privtrans] node[privlabel,near start,right] {$giveup$}	(s2)
(s1)
  edge[loop below] node[midway,below] {$proceed$} (s1)
(s2)
  edge[loop below,privtrans] node[privlabel,midway,below] {$giveup$} (s2);

%% file: Reasoning.tex


\emph{Alternating-time temporal logic} \ATLs~\short{\cite{Alur02ATL,Schobbens04ATL}}\extended{\cite{Alur97ATL,Alur02ATL,Schobbens04ATL}}
\extended{generalizes the branching-time temporal logic \CTLs\extended{~\cite{Clarke81ctl}} by replacing the path quantifiers $\Epath,\Apath$ with}%
\short{introduces} \emph{strategic modalities} $\coop{A}\gamma$, expressing that agents $A$ can enforce the temporal property $\gamma$.
\extended{While the semantics of \ATLs is typically defined for models of synchronous systems, a}\short{A} variant for asynchronous MAS was proposed recently \cite{Jamroga18por}.
We summarize the main points in this section.

\short{\para{Syntax.}}\extended{\subsection{Syntax}}
%
Let $\PV$ be a set of pro\-positional variables and $\A$ the set of all agents.
The language of \ATLs is defined as below.
\begin{center}
$\varphi::= \prop{p} \mid \neg \varphi \mid \varphi\wedge\varphi
  \mid \coop{A}\gamma$, \quad\quad
$\gamma::=\varphi \mid \neg\gamma \mid \gamma\land\gamma \mid
  \Next\gamma \mid \gamma\Until\gamma$,
\end{center}
where $\propp \in \PV$, $A \subseteq \A$, $\Next$ stands for ``next'', and $\Until$ for ``strong until''\extended{ ($\gamma_1\Until\gamma_2$ denotes that $\gamma_1$ holds until $\gamma_2$ becomes true)}.
The other Boolean operators and constants are defined as usual.
``Release'' can be defined as $\gamma_1\Release\gamma_2 \equiv \neg((\neg\gamma_1)\Until(\neg\gamma_2))$.
``Eventually'' and ``always'' can be defined as $\Sometm\gamma \equiv \true\Until\gamma$ and $\Always\gamma \equiv \false\Release\gamma$.
\extended{Moreover, the \CTLS operator ``for all paths'' can be defined as $\Apath\gamma \equiv \coop{\emptyset}\gamma$.}


\begin{example}[Conference]\label{ex:formulae}
Formula $\coop{sc}\Sometm\prop{open}$ expresses that the Steering Chair can enforce that the conference is eventually opened.
Moreover, formula $\coop{gc,oc}\Always\neg\prop{epid}$ says that the General Chair and the Organizing Chair have a joint strategy to avoid high epidemiological risk.
\end{example}
%

\short{\para{Strategies and outcomes.}}\extended{\subsection{Strategies and Outcomes}}\label{sec:strategies}
%
%
\extended{
  We adopt Schobbens' taxonomy and notation for strategy types~\cite{Schobbens04ATL}:
  \ir, \Ir, \iR, and \IR,
  where \emph{I} (resp. \emph{i}) denotes {perfect} (resp. {imperfect}) \emph{information},
  and \emph{R} (resp. \emph{r}) denotes {perfect} (resp. {imperfect}) \emph{recall}.
  In particular, an
}%
\short{An }\emph{imperfect \textbf{i}nformation/im\-perfect \textbf{r}ecall strategy (\ir-strategy) for $i$} is a function $\strat_i \colon L_i \to \events_i$ s.t. $\strat_i(l) \in \roc_i(l)$ for each $l \in L_i$.
We denote the set of such strategies by $\Sigma_i^{\ir}$.
A \emph{collective strategy} $\strat_A$ for a coalition $A = (1,\dots,m) \subseteq \A$ is a tuple of strategies, one per agent $i \in A$.
The set of $A$'s collective \ir strategies is denoted by $\Sigma_A^{\ir}$.
We will sometimes use $\strat_A(\state) = (\strat_{a_1}(\state),\dots,\strat_{a_m}(\state))$ to denote the tuple of $A$'s selections at state $\state$.

\begin{example}[Conference]\label{ex:strategy}
A collective strategy for the General Chair and the OC Chair \extended{in the conference scenario }is shown in Figure~\ref{fig:conference-strategy}.
\end{example}

An infinite sequence of global states and events $\seq = \state_0 \evt_0 \state_1 \evt_1 \state_2\dots$ is
called a\extended{n (interleaved)} \emph{path} if $\state_i \trans {\evt_i} \state_{i+1}$ for every $i \geq 0$.
$\events(\seq) = \evt_0\evt_1\evt_2\ldots$ is the sequence of events in $\seq$,
and $\seq[i] = \state_i$ is the $i$-th global state of $\seq$.
$\Pi_M(\state)$ denotes the set of all paths in model $M$ starting at $\state$.
Intuitively, the outcome of $\strat_A$ in $\state$ is the set of all the paths that can occur when the agents in $A$ follow $\strat_A$ and the agents in $\A \setminus A$ freely choose events from their repertoires. 
To define it formally, we first refine the concept of an enabled event, taking into account the choices of $A$ in strategy $\strat_A$.

\begin{definition}[Enabled events]\label{def:enabled}
Let $A = (1,\dots,m)$, $\state\in \States$, and let $\overrightarrow{\evt}_A = (\evt_1,\dots,\evt_m)$ be a tuple of events such that every $\evt_i\in \roc_i(\state^i)$.
That is, every $\evt_i$ can be selected by its respective agent $i$ at state $g$.
We say that event $\evttwo \in \events$ is \emph{enabled by $\overrightarrow{\evt}_A$ at $\state\in \States$} iff
\begin{itemize2}
\item for every $i\in Agent(\evttwo)\cap A$, we have $\evttwo=\evt_i$, and
\item for every $i\in Agent(\evttwo)\setminus A$, it holds that $\evttwo\in \roc_i(\state^i)$.
\end{itemize2}

Thus, $\evttwo$ is enabled by $\overrightarrow{\evt}_A$ if all the agents that ``own'' $\evttwo$ can choose $\evttwo$ for execution, even when $\overrightarrow{\evt}_A$ has been selected by the coalition $A$.
We denote the set of such events by $enabled(\state,\overrightarrow{\evt}_A)$. Clearly, $enabled(\state,\overrightarrow{\evt}_A) \subseteq enabled(\state)$.
\end{definition}

\begin{example}[Conference]\label{ex:conference-enabled}
Consider state $g=000$ and the choices of agents $A=\set{gc,oc}$ shown in Figure~\ref{fig:conference-strategy}, i.e., $\overrightarrow{\evt}_A=(proceed,online)$. The only events enabled by $\overrightarrow{\evt}_A$ are $proceed$ and $giveup$.
Event $onsite$ is not enabled because $A$ chose different events for execution; $online$ is not enabled because it requires synchronization which is impossible at $000$.
\end{example}

\begin{definition}[Outcome paths]\label{def:outcome}
The \emph{outcome} of strategy $\strat_A\in\Sigma_A^\ir$ in state $\state \in \States$
is the set $\outcome_M(\state,\strat_A) \subseteq \Pi_M(\state)$ such that
$\seq = \state_0 \evt_0 \state_1 \evt_1 \state_2\dots \in \outcome_M(\state,\strat_A)$
iff $\state_0 = \state$,
and $\forall i \geq 0 \quad \evt_i \in enabled(\seq[i],\strat_A(\seq[i]))$.
\end{definition}
%

One often wants to look only at paths that do not consistently ignore agents whose choice is always enabled.
Formally, a path $\pi$ satisfies \emph{concurrency-fairness} (\CF) if there is no event $\evt$ enabled
in all states of $\pi$ from $\pi[n]$ on and such that for every $\evt_i$ actually executed in $\pi[i]$, $i=n,n+1,\dots$, we have $Agent(\evt)\cap Agent(\evt_i) = \emptyset$.
We denote the set of all such paths starting at $\state$ by $\Pi_M^{\CF}(\state)$.

\begin{definition}[\CF-outcome]\label{def:cf-outcome}
The \emph{\CF-outcome} of $\strat_A\in\Sigma_A^\ir$ is defined
as $\outcome^{\CF}_M(\state,\strat_A) = \outcome_M(\state,\strat_A) \cap \Pi_M^{\CF}(\state)$.
\end{definition}

\short{\para{Strategic ability for asynchronous systems.}}\extended{\subsection{Strategic Ability for Asynchronous Systems}}\label{sec:atlsemantics}
The semantics of \ATLs[ir] in AMAS is defined by the following clause for strategic modalities~\cite{Jamroga18por}:
\begin{description2}
\item[{$\model,\state \satisf[\ir] \coop{A}\gamma$}] iff there is a strategy $\strat_A\in\Sigma_A^\ir$
s.t. $\outcome_\model(\state,\strat_A) \neq \emptyset$ and, for each path $\seq\in \outcome_\model(\state,\strat_A)$,
we have $\model,\seq \satisf[\ir] \gamma$.
\end{description2}
The clauses for Boolean and temporal operators are standard.
Moreover, the \emph{concurrency-fair semantics} 
$\satisfair[\ir]{\CF}$\extended{ of \ATL and \ATLs} is obtained
by replacing $\outcome_M(\state,\strat_A)$ with $\outcome_M^{\CF}(\state,\strat_A)$ in the above clause.

\begin{example}[Conference]\label{ex:semantics}
Clearly, formula $\coop{gc,oc}\Always\neg\prop{epid}$ holds in $(\Mconf,000)$, in both $\satisf[\ir]$ and $\satisfair[\ir]{\CF}$ semantics.
To see that, fix $\strat_{gc}(0) = proceed$ and $\strat_{gc}(1) = \strat_{oc}(0) = online$ in the collective strategy of $\set{gc,oc}$.
{Note also that $\Mconf,000 \satisf[\ir] \neg\coop{gc,oc}\Sometm\prop{closed}$ because, after executing $proceed$ and $online$ (or $onsite$), event $rest$ may be selected forever.
On the other hand, such paths are not concurrency-fair, and thus $\Mconf,000 \satisfair[\ir]{\CF} \coop{gc,oc}\Sometm\prop{closed}$.}
\end{example}

\para{Discussion.}
Strategic play assumes proactive attitude: the agents in $\coop{A}$ are free to choose \emph{any} available strategy $\strat_A$.
This is conceptually consistent with the notion of agency~\cite{Bratman87intentions}.
At the same time, it is somewhat at odds with the standard semantics of concurrent processes, where the components cannot stubbornly refuse to synchronize if that is the only way to proceed with a transition.
This seems a minor problem, but it is worrying that a strategy can have the empty set of outcomes, and equally worrying that such strategies are treated differently from the other ones.
Indeed, as we will show in the subsequent sections, the semantics proposed in~\cite{Jamroga18por} leads to a counterintuitive interpretation of strategic formulas.

%% file: Paradoxes.tex
\begin{figure}[t]
\centering
\begin{tabular}{@{}c@{}}
\begin{tikzpicture}[->,>=stealth',shorten >=1pt,auto,node distance=2.1cm,transform shape,semithick,scale=\scale]
\input{conference-2-unfolding+strat.tex}
\end{tikzpicture}
\end{tabular}
\caption{Model $\Mconf$ for the conference scenario. We highlight the transitions enabled by the strategy in Figure~\ref{fig:conference-strategy}, and the resulting reachable states. }
\label{fig:conference-model}
\end{figure}
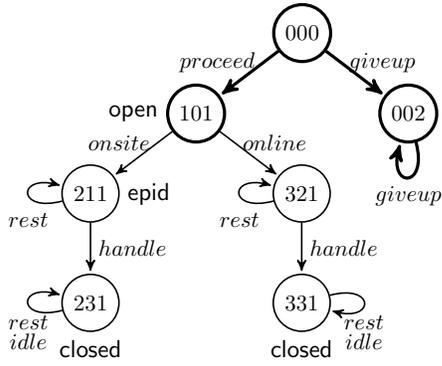

Starting with this section, we describe some problematic phenomena that follow from the straightforward combination of strategic ability with models of concurrent systems, proposed in~\cite{Jamroga18por}. 
We also show how to extend the representations and modify their execution semantics to avoid the counterintuitive interpretation of formulas. 

\subsection{Deadlock Strategies and Finite Paths}\label{sec:deadlocks}


An automata network is typically required to produce no deadlock states, i.e., every global state in its {composition} must have at least one outgoing transition. Then, all the maximal paths are infinite, and it is natural to refer to only infinite paths in the semantics of temporal operators.
In case of AMAS, the situation is more delicate. Even if the AMAS as a whole produces no deadlocks, some strategies might, which makes the interpretation of strategic modalities cumbersome. We illustrate this on the following example.

\begin{example}[Conference]\label{ex:finitepaths}
Recall the 3-agent AMAS of Figure~\ref{fig:conference}, together with its model $\Mconf$ (Figure~\ref{fig:conference-model}).
Clearly, $\Mconf$ has no deadlock states.
Let us now look at the collective strategies of coalition $\set{gc,oc}$, with agent $sc$ serving as the opponent.
It is easy to see that the coalition has no way to prevent the opening of the conference, i.e., it cannot prevent the system from reaching state $101$.
However, the strategy depicted in Figure~\ref{fig:conference-strategy} produces only one \emph{infinite} path: $(000 \,giveup\, 002 \,giveup\, \dots)$.
Since the \ATLs semantics in Section~\ref{sec:atl} disregards finite paths, we get that $\Mconf,000 \models \coop{gc,oc}\Always\neg\prop{open}$, which is counterintuitive.
\end{example}

Things can get even trickier.
{In particular, the outcome of a strategy can be empty -- in fact, it may even happen that a coalition has only strategies with empty outcomes.}

\begin{figure}[t]
\centering
\begin{tabular}{@{}cc@{}}
	\begin{tabular}{@{}c@{}}
  	\begin{tikzpicture}[->,>=stealth',shorten >=1pt,auto,node distance=2.1cm,transform shape,semithick,scale=\scale]
       \input{voter-simple.tex}
    \end{tikzpicture}
    \end{tabular}
 &
	\begin{tabular}{@{}c@{}}
    \begin{tikzpicture}[->,>=stealth',shorten >=1pt,auto,node distance=2.1cm,transform shape,semithick,scale=\scale]
     \input{ebm.tex}
    \end{tikzpicture}
    \end{tabular}
\end{tabular}
\caption{Casting a ballot: voter $v$ (left) and EBM $ebm$ (right)}
\label{fig:voting}
\end{figure}
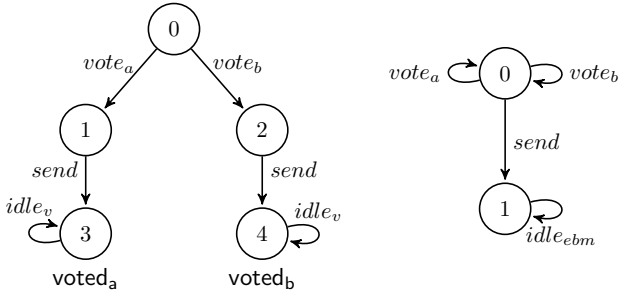

\begin{example}[Voting]\label{ex:voting}\label{ex:emptyoutcome}
Consider the AMAS in Figure~\ref{fig:voting} that depicts a simple voting scenario.
A voter $v$ can fill in an electronic ballot with a vote for candidate $\prop{a}$ or $\prop{b}$, and then push the $send$ button.
The Electronic Ballot Machine $ebm$ duly registers the choices of the voter.
%
Note that all the \emph{joint} strategies of $\set{v,ebm}$ produce only finite \short{runs}\extended{sequences of transitions}.
This is because $ebm$ must choose a single event at location $0$ in a memoryless strategy, and thus $v$ and $ebm$ are bound to ``miscoordinate'' either at the first or at the second step.
Since finite paths are not included in the outcome sets, and the semantics in Section~\ref{sec:atlsemantics} rules out strategies with empty outcomes, we get that $IIS(\AMASVoting),00 \satisf \neg\coop{v,ebm}\Sometm\top$, which is quite strange.
\end{example}

Notice that removing the non-emptiness requirement from the semantic clause in Section~\ref{sec:atlsemantics} does not help.
In that case, any joint strategy of $\set{v,ebm}$ could be used to demonstrate that $\coop{v,ebm}\Always\bot$.

\subsection{Solution: Adding Silent Transitions}\label{sec:adding-epsilon}

To deal with the problem, we augment the model of the system with special ``silent'' transitions, labeled by $\epsilon$, that are fired whenever no ``real'' transition can occur.
In our case, the $\epsilon$-transitions account for the possibility that some agents miscoordinate and thus block the system.
Moreover, we redefine the outcome set of a strategy so that an $\epsilon$-transition is taken whenever such miscoordination occurs.

\begin{definition}[Undeadlocked IIS]\label{def:undeadlockedIIS}
Let $\AMAS$ be an AMAS, and assume that no agent in \AMAS has $\epsilon$ in its alphabet of events.
The \emph{undeadlocked model of \AMAS}, denoted $\modelEpsilon = \IISEps(\AMAS)$, extends the model $\model = IIS(\AMAS)$ as follows:
\begin{itemize2}
\item $\events_{\modelEpsilon} = \events_{\model} \cup \set{\epsilon}$, where $Agent(\epsilon) = \emptyset$;\
\item For each $\state\in\States$, we add the transition $\state \trans\epsilon \state$ iff there is a selection of agents' choices $\overrightarrow{\evt}_A = (\evt_1,\dots,\evt_k)$, $\evt_i\in \roc_i(g^i)$, such that $enabled_{\model}(\state,\overrightarrow{\evt}_A) = \emptyset$.
Then, for every $A \in \A$, we also fix $enabled_{\modelEpsilon}(\state,\overrightarrow{\evt}_A) = enabled_{\modelEpsilon}(\state,\overrightarrow{\evt}_A) \cup \set{\epsilon}$.
\end{itemize2}
In other words, ``silent'' loops are added in the states where a combination of the agents' actions can block the system.
\end{definition}

\noindent
Paths are defined as in Section~\ref{sec:iis}.
The following is trivial.

\begin{smalltheorem}\label{prop:nonempty}
For any AMAS $\AMAS$, any state $\state\in \IISEps(\AMAS)$, and any strategy $\strat_A$, we have that $enabled_{\IISEps(\AMAS)}(\state,\strat_A(state)) \neq \emptyset$.
\end{smalltheorem}

\begin{example}[Conference]\label{ex:conference-undeadlocked}\label{ex:conference-undeadlocked-paths}
The undeadlocked model $\Mconf^\eps$ of the conference scenario (Example~\ref{ex:conference-amas}) extends the model in Figure~\ref{fig:conference-model} with one $\eps$-loop at state $101$. The loop models the situation when the agents choose $(onsite, online, proceed)$ or $(online, onsite, proceed)$.
We leave it for the reader to check that, at the other states, all the combinations of choices enable at least one transition.

For the strategy in Example~\ref{ex:finitepaths}, notice that its outcome in $\Mconf^\eps$ contains \emph{two} infinite paths: not only $(000 \,giveup\, 002 \,giveup\, 002 \dots)$, but also $(000 \,proceed\, 101 \,\epsilon\, 101 \dots)$.
Since the latter path invalidates \extended{the temporal formula }$\Always\neg\prop{open}$, we get that $\Mconf,000 \not\models \coop{gc,oc}\Always\neg\prop{open}$, as expected.
\end{example}

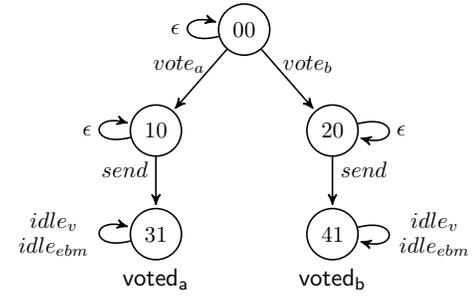
\begin{figure}[t]
\centering
  	\begin{tikzpicture}[->,>=stealth',shorten >=1pt,auto,node distance=2.1cm,transform shape,semithick,scale=\scale]
       \input{voter+ebm-undeadlocked.tex}
    \end{tikzpicture}
\caption{Undeadlocked IIS for the voting scenario}
\label{fig:voting-undeadlocked}
\end{figure}

\begin{example}[Voting]\label{ex:voting-undeadlocked}
The undeadlocked model for the voting scenario is presented in Figure~\ref{fig:voting-undeadlocked}.
Note that formula $\neg\coop{v,ebm}\Sometm\top$ does not hold anymore, because the joint strategies of $\set{v,ebm}$ have nonempty outcomes in $\IISEps(\AMASVoting)$.
On the other hand, the formula $\coop{v}\Sometm\prop{voted_{a}}$ (and even $\coop{v,ebm}\Sometm\prop{voted_{a}}$) does not hold, which is contrary to the intuition behind the modeling.
We will come back to this issue in Section~\ref{sec:asymmetric}.
\end{example}

\para{Discussion.}
Adding ``silent'' transitions to account for the control flow when no observable event occurs is pretty standard. The crucial issue is \emph{where} to add them. Here, we add the $\eps$-transitions whenever a subset of agents might {choose} to miscoordinate (and stick to their choices). Again, this is in line with the notion of agency and strategic play in MAS~\cite{Bratman87intentions,Pauly01eff}.
In the next section, we will discuss a concept of ``agent fairness'' where the addition of $\eps$-transitions is constrained by the assumption that only a given subset of agents is fully proactive.

The examples used in this section expose an important feature of agent systems.
The execution semantics of concurrent processes is often defined by a state-transition graph (or, alternatively, by the tree of paths generated by the graph, i.e., the tree unfolding of the graph). For systems that involve proactive agents, this is not enough. Rather, the execution semantics should map from the possible coalitions and their available strategies to the outcome sets of those strategies. In this sense, the possible behaviors of an agent system should be understood via the \emph{set of possible execution trees}, rather than a single tree.
This is consistent with the theoretical model of MAS in~\cite{Goranko15patheffectivity}, based on path effectivity functions.

An alternative way out of the problem is to include finite maximal paths in the outcomes of strategies. However, the interpretation of strategic modalities over finite paths is rather nonstandard~\cite{Belardinelli18atl-finitetraces} and may pose new problems in the asynchronous setting.
Moreover, our approach allows to reuse the existing techniques and tools, which are typically built for infinite path semantics, including the verification and partial order reduction functionalities of tools like SPIN~\cite{Holzmann97spin} and STV~\cite{Kurpiewski21stv-demo}.
In general, this is a design dilemma between changing the logical semantics of the formulas vs.~updating the execution semantics of the representations. Here, we choose the latter approach.

\section{Playing Against Reactive Opponents}\label{sec:fairness}

The solution proposed in Section~\ref{sec:adding-epsilon} is based on the assumption that an agent is free to choose any event in its repertoire -- even one that prevents the system from executing anything.
The downside is that, for most systems, only safety goals can be achieved (i.e., properties specified by $\coop{A}\Always\varphi$).
For reachability, there is often a combination of the opponents' choices that blocks the execution early on, and prevents the coalition from reaching their goal.
In this section, we define a fairness-style condition that constrains the choices of more ``reactive'' opponents.
We also show a construction to verify the abilities of the coalition over the resulting paths in a technically simpler way.

\subsection{Opponent-Reactiveness}\label{sec:enfliveness}

Given a strategy $\strat_A$, the agents in $A$ are by definition assumed to be proactive.
Below, we propose an execution semantics for $\strat_A$ which assumes that  $A$ cannot be stalled forever by miscoordination on the part of the opponents.

\begin{definition}[Opponent-reactiveness]\label{def:enfliveness}\label{def:reactiveness}
A path $\seq = \state_0 \evt_0 \state_1 \evt_1 \state_2\dots$ in $\IISEps(\AMAS)$ is \emph{opponent-reactive for strategy $\strat_A$} iff we have that
$\evt_n = \epsilon$ implies $enabled(\state_n,\strat_A(\state_n)) = \set{\epsilon}$.
In other words, whenever the agents outside $A$ have a way to proceed, they must proceed.
The \emph{reactive outcome}\extended{ (or \emph{\EL-outcome})} of $\strat_A$ in $\state$, denoted $\outcomeEps(\state,\strat_A)$, is the restriction of $\outcome(\state,\strat_A)$ to its opponent-reactive paths.
%
\end{definition}

\begin{example}[Conference]\label{ex:reactiveness}
Consider the undeadlocked model $\Mconf^\eps$ of Example~\ref{ex:conference-undeadlocked}.
Path $(000 \,proceed\, 101 \,\epsilon\, 101 \dots)$ is opponent-reactive for the strategy of agents $\set{gc, oc}$ shown in Figure~\ref{fig:conference-strategy}.

On the other hand, consider coalition $\set{gc,sc}$, and the following strategy of theirs:\ $\strat_{gc}(0)=proceed, \strat_{gc}(1)=onsite, \strat_{sc}(0)=proceed$.
The same path is \emph{not} opponent-reactive for the strategy because the only opponent ($oc$) has a response at state $101$ that enables a ``real'' transition ($onsite$).
\end{example}


\begin{smalltheorem}\label{prop:epsilon}
In $\outcomeEps_{\IISEps(\AMAS)}(\state,\strat_A)$, the only possible occurrence of $\eps$ is as an infinite sequence of $\eps$-transitions following a finite prefix of ``real'' transitions.
\end{smalltheorem}
%
\begin{proof}
Take any $\seq = \state_0 \evt_0 \state_1 \evt_1 \state_2\dots \in \outcomeEps_{\IISEps(\AMAS)}(\state,\strat_A)$ such that $\epsilon$ occurs on $\seq$, and let $i$ be the first position on $\seq$ st.~$\evt_i = \epsilon$.
By Definition~\ref{def:reactiveness}, we get that $enabled(\state_i,\strat_A(\state_i)) = \set{\epsilon}$. Moreover, $state_{i+1} = \state_i$, so also $enabled(\state_{i+1},\strat_A(\state_{i+1})) = \set{\epsilon}$. Thus, $\evt_{i+1} = \epsilon$.
It follows by simple induction that $\evt_j = \epsilon$ for every $j\ge i$.
\end{proof}

The \emph{opponent-reactive semantics} $\satisfair[\ir]{\EL}$ of \ATLs is obtained
by replacing $\outcome_M(\state,\strat_A)$ with $\outcome_M^{\EL}(\state,\strat_A)$ in the semantic clause presented in Section~\ref{sec:atlsemantics}.

\subsection{Encoding Strategic Deadlock-Freeness Under Opponent-Reactiveness in AMAS}

If we adopt the assumption of opponent-reactiveness for coalition $A$, there is an alternative, technically simpler way to obtain the same semantics of strategic ability as in Section~\ref{sec:adding-epsilon}. The idea is to introduce the ``silent'' transitions already at the level of the AMAS.

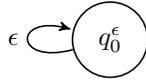
\begin{figure}[t]\centering
\begin{tikzpicture}[->,>=stealth',shorten >=1pt,auto,node distance=2.1cm,transform shape,semithick,scale=1]
 \input{epsilon-agent.tex}
\end{tikzpicture}
\caption{The auxiliary agent added in $\AMAS^\eps$}
\label{fig:undeadlocked}
\end{figure}

\begin{definition}[Undeadlocked AMAS]\label{def:undeadlockedAMAS}
The \emph{undeadlocked variant of $\AMAS$} is constructed from $\AMAS$ by adding an auxiliary agent $A_\epsilon$ with\
$L_\epsilon =\set{q_0^\epsilon}$,\
$\iota_\epsilon =q_0^\epsilon$,\
$\events_\epsilon = \set{\epsilon}$,\
$\roc_\epsilon(q_0^\epsilon) = \set{\epsilon}$,\
$T_i(q_0^\epsilon,\epsilon) = q_0^\epsilon$,\ and
$\PV_\epsilon = \emptyset$.
In other words, we add a module with a single local state and a ``silent'' loop labeled by $\epsilon$, as in Figure~\ref{fig:undeadlocked}.
We will denote the undeadlocked variant of $\AMAS$ by $\epsAMAS$.
Note that $\epsAMAS$ can be seen as a special case of AMAS. Thus, the outcome sets and reactive outcomes of strategies in $IIS(\epsAMAS)$ are defined exactly as before.
\end{definition}

\extended{
  \begin{example}[Voting]\label{ex:ASV-undeadlocked}
  The undeadlocked AMAS $\ASV{1}{2}^\epsilon$ is obtained by augmenting $\ASV{1}{2}$ with the auxiliary agent in Figure~\ref{fig:undeadlocked}.
  \end{example}
}

Obviously, the extra agent adds $\eps$-loops to the model of $\AMAS$, i.e., to $IIS(\AMAS)$.
We show now that, under the assumption of opponent-reactiveness, the view of $A$'s strategic ability in the undeadlocked AMAS $\AMAS^\eps$ corresponds precisely to $A$'s abilities in the undeadlocked model of the original AMAS $\AMAS$, i.e., $\IISEps(\AMAS)$.
This allows to deal with deadlocks and finite paths without redefining the execution semantics for AMAS, set in Definition~\ref{def:canonical-iis},
and thus use the existing tools such as SPIN~\cite{Holzmann97spin} in a straightforward way.

\begin{smalltheorem}\label{prop:epsilon2}
Let $A\subseteq\Agt$.
In $\outcomeEps_{IIS(\epsAMAS)}(\state,\strat_A)$,
the only possible occurrence of $\eps$ is as an infinite suffix of $\eps$-transitions.
\end{smalltheorem}
\begin{proof}
Analogous to Proposition~\ref{prop:epsilon}.
\end{proof}

\begin{theorem}\label{prop:undeadlocked}
For every strategy $\strat_A$ in $\AMAS$, we have that
\[\outcomeEps_{\IISEps(\AMAS)} (\state,\strat_A) = \outcomeEps_{IIS(\epsAMAS)}(\state,\strat_A).\]
\end{theorem}
\begin{proof}
%
{$\pmb{\outcomeEps_{\IISEps(\AMAS)} (\state,\strat_A) \subseteq \outcomeEps_{IIS(\epsAMAS)}(\state,\strat_A)}$}:\
Consider any $\seq = \state_0 \evt_0 \state_1 \evt_1 \state_2\dots \in \outcomeEps_{\IISEps(\AMAS)}(\state,\strat_A)$. 
If there are no $\epsilon$-transitions on $\seq$, we have that $\seq \in \outcomeEps_{IIS(\AMAS)}(\state,\strat_A) \subseteq \outcomeEps_{IIS(\epsAMAS)}(\state,\strat_A)$, QED.
Suppose that $\seq$ includes $\epsilon$-transitions, with $\evt_i$ being the first one. 
  Then, we have that $\evt_j\neq\epsilon$ and $\evt_j \in enabled_{\IISEps(\AMAS)}(\state_j,\strat_A(\state_j))$ for every $j<i$,
	hence also $\evt_j \in enabled_{IIS(\AMAS)}(\state_j,\strat_A(\state_j)) \subseteq enabled_{IIS(\epsAMAS)}(\state_j,\strat_A(\state_j))$. \textbf{(*)} \\
  By Proposition~\ref{prop:epsilon}, $\state_j = \state_i$ and $\evt_j = \epsilon$ for every $j\ge i$.
  By Definition~\ref{def:reactiveness}, $enabled_{\IISEps(\AMAS)}(\state_j,\strat_A(\state_j)) = \set{\epsilon}$. Hence, $enabled_{IIS(\AMAS)}(\state_j,\strat_A(\state_j)) = \emptyset$ and $enabled_{IIS(\epsAMAS)}(\state_j,\strat_A(\state_j)) = \set{\epsilon}$. \textbf{(**)} \\
  Thus, by \textbf{(*)} and \textbf{(**)}, $\seq \in \outcomeEps_{IIS(\epsAMAS)}(\state,\strat_A)$, QED.

\smallskip\noindent
{$\pmb{\outcomeEps_{IIS(\epsAMAS)} (\state,\strat_A) \subseteq \outcomeEps_{\IISEps(\AMAS)}(\state,\strat_A)}$}:\
Analogous, with Proposition~\ref{prop:epsilon2} used instead of Proposition~\ref{prop:epsilon}.
\end{proof}

\para{Discussion.}
Opponent-reactiveness is to strategic properties what fairness conditions are to temporal properties of asynchronous systems.
If an important property cannot be satisfied in all possible executions, it may at least hold under some reasonable assumptions about which events can be selected by whom in response to what.
Clearly, the condition can be considered intuitive by some and problematic by others. The main point is, unlike in the previous semantics, now it is made explicit, and can be adopted or rejected depending on the intuition. \extended{Note that the semantic extensions proposed in this paper (silent transitions and nondeterministic choices for strategies) make sense both with and without opponent-reactiveness.}

Note that, under the reactiveness assumption, we have that $\Mconf^\eps,000 \satisfair[\ir]{\EL} \coop{gc,sc}\Sometm\prop{epid}$\ and\ $\Mconf^\eps,000 \satisfair[\ir]{\EL} \coop{oc}\Always\neg\prop{epid}$.
This seems to contradict the commonly accepted requirement of \emph{regularity} in games~\cite{Pauly01phd}.
However, the contradiction is only superficial, as the two formulas are evaluated \emph{under different execution assumptions}: for the former, we assume agent $oc$ to be reactive, whereas the latter assumes $gc$ and $sc$ to react to the strategy of $oc$.

\section{Concurrency-Fairness Revisited}

In Def.~\ref{def:cf-outcome}, we recalled the notion of concurrency-fair outcome of~\cite{Jamroga18por}.
The idea was to remove from $out(\state,\strat_A)$ the paths that consistently ignore agents whose events are enabled \emph{at the level of the whole model}.
Unfortunately, the definition has unwelcome side effects, too.

\subsection{Problems with Concurrency-Fairness}\label{sec:problemsCF}

We first show that, contrary to intuition, Definition~\ref{def:cf-outcome} automatically disregards \emph{deadlock paths}, i.e., paths with finitely many ``real'' transitions.

\begin{smalltheorem}\label{prop:finitepathsCF}
Consider an AMAS $\AMAS$ and a path $\seq$ in $\IISEps(\AMAS)$ such that, from some point $i$ on, $\seq$ includes only $\epsilon$-transitions.
Then, for every strategy $\strat_A$ in $\AMAS$, we have that $\seq\notin\outcomeCF_{\IISEps(\AMAS)}(\state,\strat_A)$.
\end{smalltheorem}
\begin{proof}
Take $\seq$ as above, i.e., $\seq = \state_0 \evt_0 \state_1 \evt_1 \dots \state_i\epsilon\state_i\epsilon\state_i\dots$.
Since the transition function in $\IISEps(\AMAS)$ is serial, there must be some event $\evttwo\neq\epsilon$ enabled in $\state_i$.
In consequence, $\evttwo$ is always enabled from $i$ on, but none of its ``owners'' in $Agent(\evttwo)$ executes an event on $\seq$ after $i$.
Hence, $\seq$ does not satisfy \CF, and does not belong to $\outcomeCF_{\IISEps(\AMAS)}(\state,\strat_A)$ for any strategy $\strat_A$.
\end{proof}

Thus, the \CF condition eliminates all the deadlock paths from the outcome of a strategy (for instance, the path $(000 \,proceed\, 101 \,\epsilon\, 101 \dots)$ in Example~\ref{ex:conference-undeadlocked-paths}).
In consequence, reasoning about concurrency-fair paths suffers from the problems that we identified in Section~\ref{sec:deadlocks}, even for undeadlocked models.
Moreover, combining the temporal and strategic fairness (i.e., \CF and \EL) collapses the undeadlocked execution semantics altogether, see below.

\begin{smalltheorem}\label{prop:ELplusCF}
Reasoning about reactive \emph{and} fair outcomes in an undeadlocked model reduces to reasoning about the fair executions in the original model without $\eps$-transitions.
Formally, let $\outcomeEpsCF_{M}(\state,\strat_A) = \outcomeEps_{M}(\state,\strat_A) \cap \outcomeCF_{M}(\state,\strat_A)$.
For any AMAS $\AMAS$ and any strategy $\strat_A$ in $\AMAS$, we have:

\smallskip
\centerline{$\outcomeEpsCF_{\IISEps(\AMAS)}(\state,\strat_A) = \outcomeCF_{IIS(\AMAS)}(\state,\strat_A)$.}
\end{smalltheorem}
\begin{proof}
Clearly, we have $\outcomeCF_{IIS(\AMAS)}(\state,\strat_A) \subseteq \outcomeEpsCF_{\IISEps(\AMAS)}(\state,\strat_A)$,
since $\outcomeEpsCF_{\IISEps(\AMAS)}(\state,\strat_A)$ can only add to $\outcomeCF_{IIS(\AMAS)}(\state,\strat_A)$ new paths that include $\epsilon$-transitions.

For the other direction, take any $\seq\in\outcomeEpsCF_{\IISEps(\AMAS)}(\state,\strat_A)$, and suppose that it contains an $\epsilon$-transition.
By Proposition~\ref{prop:epsilon}, it must have an infinite suffix consisting only of $\epsilon$-transitions.
Then, by Proposition~\ref{prop:finitepathsCF}, $\seq\notin\outcomeCF_{\IISEps(\AMAS)}(\state,\strat_A)$, which leads to a contradiction.
Thus, $\seq$ contains only transitions from $IIS(\AMAS)$, and hence $\seq\in\outcomeCF_{IIS(\AMAS)}(\state,\strat_A)$, QED.
\end{proof}

\subsection{Strategic Concurrency-Fairness}

So, how should fair paths be properly defined for strategic reasoning? The answer is simple: in relation to the outcome of the strategy being executed.

\begin{definition}[Strategic \textbf{CF}]\label{def:scf-outcome}
$\seq = \state_0 \evt_0 \state_1 \evt_1 \state_2\dots$ is a \emph{concurrency-fair path for strategy $\strat_A$ and state $\state$}
iff $\state_0 = \state$, and there is no event $\evt$ s.t., for some $n$ and all $i\ge n$, we have $\evt \in enabled(\seq[i],\strat_A(\seq[i]))$ and $Agent(\evt)\cap Agent(\evt_i) = \emptyset$.
That is, agents with an event always enabled \emph{by $\strat_A$} cannot be ignored forever.

The \emph{\SCF-outcome} of $\strat_A\in\Sigma_A^\ir$ is defined
as $\outcomeSCF_M(\state,\strat_A) = \{\seq\in\outcome_M(\state,\strat_A) \mid \seq\text{ is concurrency-fair for }\strat_A,\state\}$.
\end{definition}

{
The following formal results show that \SCF does not suffer from the problems demonstrated in Section~\ref{sec:problemsCF}.

\begin{smalltheorem}\label{prop:finitepathsSCF}
There is an AMAS $\AMAS$, a strategy $\strat_A$ in $\AMAS$, and a deadlock path $\seq$ in $\IISEps(\AMAS)$ such that $\seq$ is concurrency-fair for $\strat_A$.
\end{smalltheorem}
\begin{proof}
To demonstrate the property, it suffices to take the AMAS and the strategy of $\set{gc,oc}$ depicted in Figure~\ref{fig:conference}, and the path $\seq=(000 \,proceed\, 101 \,\epsilon\, 101 \dots)$.
\end{proof}
}

\begin{theorem}\label{prop:react-vs-scf}
Opponent-reactiveness and strategic \extended{concurrency-fairness}\short{\CF} are incomparable.
Formally, there exists an AMAS $\AMAS$, a state $\state$ in $\IISEps(\AMAS)$, and a strategy $\strat_A$ such that
$\outcomeSCF_{\IISEps(\AMAS)}(\state,\strat_A) \not\subseteq \outcome^{\EL}_{\IISEps(\AMAS)}(\state,\strat_A)$, and vice versa.
\end{theorem}
\begin{proof}
Consider the undeadlocked model $\Mconf^\eps$ in Example~\ref{ex:conference-undeadlocked}, and the strategy discussed in Example~\ref{ex:reactiveness}:
$\strat_{gc}(0)=proceed$,~$\strat_{gc}(1)=onsite$,~$\strat_{sc}(0)=proceed$.
Let $\pi_1 = (000 \,proceed\, 101 \,\epsilon\, 101  \, onsite\, 211 \, rest\, 211 \linebreak handle\, 211 \, rest\, 211\dots)$.
We have $\pi_1 \in \outcomeSCF_{\Mconf^\eps}(\state,\strat_A)$, but $\pi_1 \notin \outcome^{\EL}_{\Mconf^\eps}(\state,\strat_A)$.
On the other hand,
for path $\pi_2 = (000 \,proceed\, 101 \,onsite\, 211 \,rest\, 211 \,rest\, \dots)$, we have that
$\pi_2 \notin \outcomeSCF_{\Mconf^\eps}(\state,\strat_A)$, but $\pi_2 \in \outcome^{\EL}_{\Mconf^\eps}(\state,\strat_A)$.
\end{proof}

\para{Discussion.}
Theorem~\ref{prop:react-vs-scf} suggests that reactiveness and fairness conditions arise from orthogonal concerns.
The two concepts refer to different factors that influence which sequences of events can occur.
Opponent-reactiveness constrains the choices that (a subset of) the agents can select.
Concurrency-fairness and its strategic variant restrict the way in which the ``scheduler'' (Nature, Chance, God...) can choose from the events selected by the agents.



\section{\mbox{Strategies in Asymmetric Interaction}}\label{sec:asymmetric}

Now, we point out that AMAS are too restricted to model the strategic aspects of asymmetric synchronization in a natural way (e.g., a sender sending a message to a receiver).

\subsection{Simple Choices are Not Enough}

We demonstrate the problem on an example.

{
\begin{example}[Voting]\label{ex:paradox2}
As already pointed out, we have $\IISEps(\AMASVoting),00 \not\satisf \coop{v,ebm}\Sometm\prop{voted_{a}}$ in the model of Example~\ref{ex:voting}.
This is because receiving a vote for $\prop{a}$, a vote for \prop{b}, and the signal to send the vote, belong to \emph{different choices} in the repertoire of the EBM, and the agent can only select one of them in a memoryless strategy.
Moreover, formula $\coop{ebm}\Sometm\prop{voted_{a}}$ holds under the condition of opponent-reactiveness, i.e., the EBM can force a reactive voter to vote for a selected candidate.
Clearly, it was not the intention behind the AMAS: the EBM is supposed to \emph{listen} to the choice of the voter. No matter whose strategies are considered, and who reacts to whose actions, the EBM should have no influence on what the voter votes for.
\end{example}
}

The problem arises because the repertoire functions in AMAS are based on the assumption that the agent can choose any single event in $\roc_i(l_i)$.
This does not allow for natural specification of situations when the exact transition is determined by another agent.
{For the AMAS in Example~\ref{ex:voting}, the decision to vote for candidate \prop{a} or \prop{b} (or to press $send$) should belong solely to the voter.
Thus, setting the EBM repertoire as $\roc_{ebm}(0) = \set{vote_a, vote_b, send}$ does not produce a good model of strategic play in the scenario.}

\subsection{AMAS with Explicit Control}

As a remedy, we extend the representations so that one can indicate which agent(s) control the choice between events.

\begin{definition}[AMAS with explicit control]\label{def:extendedAMAS}
Everything is exactly as in Definition~\ref{def:amas}, except for the repertoires of choices, which are now functions
$\roc_i: L_i \to 2^{2^{\events_i}\setminus\{\emptyset\}}\setminus\{\emptyset\}$.
That is, $\roc_i(l)$ lists nonempty subsets of events $X_1, X_2, \dots \subseteq \events_i$, each capturing an available choice of $i$ at the local state $l$.
If the agent chooses $X_j = \set{\evt_1,\evt_2,\dots}$, then only an event in that set can be executed
within the agent's module; however, the agent has no firmer control over which one will be fired.
Accordingly, we assume that $T_i(l,\evt)$ is defined iff $\evt\in \bigcup \roc_i(l)$.\footnote{
  For a set of sets $X$, we use $\bigcup X$ to denote\extended{ its ``flattening''} $\bigcup_{x\in X}x$. }
\end{definition}

Notice that the AMAS of Definition~\ref{def:amas} can be seen as a special case where $\roc_i(l)$ is always a list of singletons.
The definitions of IIS and undeadlocked IIS stay the same, as agents' repertoires of choices are not actually used to generate the state-transition structure for the model of $\AMAS$.
Moreover, undeadlocked AMAS with explicit control can be obtained analogously to Definition~\ref{def:undeadlockedAMAS} by adding the auxiliary ``epsilon''-agent with $\roc_\epsilon(q_0^\epsilon) = \set{\set{\epsilon}}$ in its sole local state.

Strategies still assign choices to local states; hence, the type of agent $i$'s strategies
is now $\strat_i \colon L_i \to 2^{\events_i}\setminus\{\emptyset\}$ s.t. $\strat_i(l) \in \roc_i(l)$.
The definition of the outcome set is updated accordingly, see below.
\begin{definition}[Outcome sets for AMAS with explicit control]\label{def:outcome3}
First, we lift the set of events enabled by $\overrightarrow{\evt}_A = (\evt_1,\dots,\evt_m)$ at $\state$ to match the new type of repertoires and strategies.
Formally, $\evttwo\in enabled(\state,\overrightarrow{\evt}_A)$ iff:\
(1) for every $i\in Agent(\evttwo)\cap A$, we have $\evttwo\in\evt_i$, and\
(2) for every $i\in Agent(\evttwo)\setminus A$, it holds that $\evttwo\in\bigcup \roc_i(\state^i)$.

The {outcome}, {\EL-outcome}, and \SCF-outcome of $\strat_A$ in $M,\state$
are given as in Definitions~\ref{def:outcome}, \ref{def:enfliveness}, and~\ref{def:scf-outcome}.
\end{definition}

{
\begin{example}[Voting]\label{ex:ASV-extprotocols}
We improve our voting model by assuming repertoires of choices for the voter and the EBM as follows:
$\roc_{ebm}(0) = \set{\set{vote_a,vote_b,send}}$,
$\roc_v(0) = \set{\set{vote_a},\set{vote_b}}$,
$\roc_v(1) = \roc_v(2) = \set{\set{send}}$, etc.
That is, the voter's choices are as before, but the EBM only listens to what the voter selects.

Clearly, $\coop{v,ebm}\Sometm\prop{voted_{a}}$ holds in the new AMAS.
Moreover, $\coop{ebm}\Sometm\prop{voted_{a}}$ does not hold anymore, even assuming opponent-reactiveness.
\end{example}
}

It is easy to see that Propositions~\ref{prop:nonempty}, \ref{prop:epsilon}, \ref{prop:epsilon2}, and~\ref{prop:finitepathsSCF}, as well as Theorems~\ref{prop:undeadlocked} and~\ref{prop:react-vs-scf} still hold in AMAS with explicit control.

\para{Discussion.}
%
When reasoning about strategic play of asynchronous agents, two kinds of asymmetry come into the picture.
On the one hand, the processes (agents) being modeled often synchronize in an asymmetric way.
For example, the sender chooses which message to send to the receiver.
On the other hand, the agents $A$ in formula $\coop{A}\varphi$ choose the strategy and thus push the other agents to respond accordingly.
The variant of AMAS introduced in~\cite{Jamroga18por} does not allow to capture the former kind of asymmetry.
In consequence, the choice between the available synchronization branches belongs solely to the agents indicated by the formula.
Unfortunately, there is no natural way to model the converse situation, i.e., when the agents in $\coop{A}$ are forced by the choices of their opponents.
With the new variant of AMAS, we extend the representations so that the modeler can explicitly specify the degree of autonomy of each participating agent. Without that, the degree of autonomy is implicit and comes from the formula being evaluated.

\extended{\para{Related modeling approaches.}}
Various forms of asymmetric synchronization are present in most process algebras. For example, $\pi$-calculus distinguishes between the action $\overline{c}\langle a\rangle$ of sending the value $a$ on channel $c$, and action $c(x)$ of listening on channel $c$ and storing whatever comes in variable $x$.
CSP goes further, and allows for a similar degree of flexibility to ours through suitable combinations of deterministic choice, nondeterministic choice, and interface parallel operators.
Other synchronization primitives are also possible, see e.g.~\cite{Bloem15paramVerif}\extended{ for an overview}.
Instead of allowing for multiple synchronization primitives, we come up with a single general primitive that can be instantiated to cover different kinds of interaction.

We note in passing the similarity of our new repertoire functions in Definition~\ref{def:extendedAMAS} to state effectivity functions~\cite{Pauly01eff,Pauly02modal} and especially alternating transition systems~\cite{Alur98ATL}.


%% file: conference-2-unfolding+strat.tex
\tikzstyle{state}=[circle,fill=none,draw=black,text=black,minimum size=0.8cm]
\tikzstyle{reachstate}=[state,very thick]
\tikzstyle{strat}=[very thick]

\node[reachstate] (s000) {$000$}; 
\node[reachstate] (s101) [below left=0.6cm and 1cm of s000, label=left:{\large $\prop{open}$}] {$101$};
\node[reachstate] (s002) [below right=0.6cm and 1cm of s000] {$002$};
\node[state] (s211) [below left=0.6cm and 1cm of s101, label=right:{\large $\prop{epid}$}] {$211$};
\node[state] (s321) [below right=0.6cm and 1cm of s101] {$321$};
\node[state] (s231) [below=0.8cm of s211, label=below:{\large $\prop{closed}$}] {$231$};
\node[state] (s331) [below=0.8cm of s321, label=below:{\large $\prop{closed}$}] {$331$};

\path
(s000)
  edge[strat] node[near start, left] {$proceed$}	(s101)
  edge[strat] node[near start, right] {$giveup$} (s002)
(s101)
  edge node[near start, left] {$onsite$}	(s211)
  edge node[near start, right] {$online$}	(s321)
(s002)
  edge[strat,loop below] node[midway,below]{$giveup$} (s002)
(s211)
  edge [loop left] node[midway,below=5pt]{$rest$} (s211)
  edge node[midway,right] {$handle$}	(s231)
(s321)
  edge [loop left] node[midway,below=5pt]{$rest$} (s321)
  edge node[midway,right] {$handle$}	(s331)
(s231)
  edge [loop left] node[midway,below=2pt]{$rest$} node[midway,below=10pt]{$idle$} (s231)
(s331)
  edge [loop right] node[midway,below=2pt]{$rest$} node[midway,below=10pt]{$idle$} (s331);

%% file: voter-simple.tex
\tikzstyle{every state}=[fill=none,draw=black,text=black,minimum size=0.8cm]
\tikzstyle{privlabel}=[text=\private]
\tikzstyle{privtrans}=[draw=\private]

\node[state] (s0) {$0$}; 
\node[state] (s1) [below left=1cm and 0.8cm of s0] {$1$};
\node[state] (s2) [below right=1cm and 0.8cm of s0] {$2$};
\node[state] (s3) [below=0.8cm of s1, label=below:{\large $\prop{voted_a}$}] {$3$};
\node[state] (s4) [below=0.8cm of s2, label=below:{\large $\prop{voted_b}$}] {$4$};

\path
(s0)
  edge node[near start,left] {$vote_a$} (s1)
  edge node[near start,right] {$vote_b$}	(s2)
(s1)
  edge node[near start,left] {$send$} (s3)
(s2)
  edge node[near start,right] {$send$} (s4)
(s3)
  edge [privtrans,loop left] node[privlabel,midway,above=4pt] {$idle_v$} (s3)
(s4)
  edge [privtrans,loop right] node[privlabel,midway,above=2pt] {$idle_v$} (s4);

%% file: ebm.tex
\tikzstyle{every state}=[fill=none,draw=black,text=black,minimum size=0.8cm]
\tikzstyle{privlabel}=[text=\private]
\tikzstyle{privtrans}=[draw=\private]

\node[state] (s0) {$0$}; 
\node[state] (s1) [below=1.3cm of s0] {$1$};

\path
(s0)
  edge [loop left] node[midway, left] {$vote_a$} (s0)
  edge [loop right] node[midway, right] {$vote_b$} (s0)
  edge  node[midway,right] {$send$} (s1)
(s1)
  edge [privtrans,loop right] node[privlabel,midway,below=4pt] {$idle_{ebm}$} (s1);

%% file: voter+ebm-undeadlocked.tex
\tikzstyle{every state}=[fill=none,draw=black,text=black,minimum size=0.8cm]
\tikzstyle{privlabel}=[text=\private]
\tikzstyle{privtrans}=[draw=\private]

\node[state] (s00) {$00$}; 
\node[state] (s10) [below left=1cm and 0.8cm of s00] {$10$};
\node[state] (s20) [below right=1cm and 0.8cm of s00] {$20$};
\node[state] (s31) [below=0.8cm of s10, label=below:{\large $\prop{voted_a}$}] {$31$};
\node[state] (s41) [below=0.8cm of s20, label=below:{\large $\prop{voted_b}$}] {$41$};

\path
(s00)
  edge [loop left] node[midway, left] {$\epsilon$} (s00)
  edge node[near start,left] {$vote_a$} (s10)
  edge node[near start,right] {$vote_b$}	(s20)
(s10)
  edge [loop left] node[midway, left] {$\epsilon$} (s10)
  edge node[near start,left] {$send$} (s31)
(s20)
  edge [loop right] node[midway, right] {$\epsilon$} (s20)
  edge node[near start,right] {$send$} (s41)
(s31)
  edge [privtrans,loop left] node[privlabel,midway,left] {\LARGE $\genfrac{}{}{0pt}{}{idle_v}{idle_{ebm}}$} (s31)
(s41)
  edge [privtrans,loop right] node[privlabel,midway,right] {\LARGE $\genfrac{}{}{0pt}{}{idle_v}{idle_{ebm}}$} (s41);

%% file: epsilon-agent.tex
  \tikzstyle{every state}=[fill=none,draw=black,text=black,minimum size=1cm]

  \node[state] (s0) {$q_0^\epsilon$}; 

  \path (s0)	edge [loop left]	node 	{$\epsilon$}	(s0);

%% file: Po-reduction-short.tex
\emph{Partial order reduction (POR)} has been defined for temporal and temporal-epistemic logics
without ``next''~\short{\cite{peled-representatives,GKPP99,LomuscioPQ10b}}%
\extended{\cite{peled-representatives,PenczekSGK00,GKPP99,LomuscioPQ10b}},
and recently extended to strategic specifications~\cite{Jamroga18por}.
The idea is {to take a network of automata (AMAS in our case), and use depth-first search through the space of global states to} generate a reduced model that satisfies exactly the same formulas as the full model.
%
Essentially, POR removes paths that change only the interleaving order of an ``irrelevant'' event with another event.
Importantly, the method generates the reduced model directly from the representation, without generating the full model at all.

{
\subsection{Correctness of POR in the New Semantics}

POR is a powerful technique to contain state-space explosion and facilitate verification, cf.~e.g.~the experimental results in~\cite{Jamroga20POR-JAIR}.
In this paper, we extend the class of models, and modify their execution semantics.
We need to show that the reduction algorithm in~\cite{Jamroga18por}, defined for the flawed semantics of ability, is still correct after the modifications. 
Our main technical result in this respect is Theorem~\short{7.1}\extended{\ref{prop:ae-atlir-corollary}}, presented below.
The detailed definitions, algorithms and proofs are technical (and rather tedious) adaptations of those in~\cite{Jamroga18por}.
We omit them here for lack of space, and refer the inquisitive reader to \short{the extended version of the paper~\cite{Jamroga20paradoxes-tr}}\extended{Appendix~\ref{sec:por-details}}.
}

\bigskip\noindent
\textbf{Theorem~\short{7.1}\extended{\ref{prop:ae-atlir-corollary}}.}
Let $\modelEps=\fullmodel(\epsAMAS)$, $\modelEpsilon=\IISEps(\AMAS)$ and let $A\subseteq\Agt$ be a subset of agents. Moreover, let $\modelEpsPrim \subseteq \modelEps$ and {$\modelEpsilon{'} \subseteq \modelEpsilon$} be the reduced models generated by DFS with the choice of {enabled events $E(g')$} given by conditions {\bf C1, C2, C3} and the independence relation $I_{A,\hatPV}$.
For each \sATLS[\ir] formula $\varphi$ over $\hatPV$, that refers only to coalitions $\hat{A}\subseteq A$,
we have:
\begin{enumerate}
\item $\modelEps,\iota \satisfair[\ir]{\EL} \varphi$\quad iff \quad $\modelEpsPrim,\iota' \satisfair[\ir]{\EL} \varphi$,\quad and
\item $\modelEpsilon,\iota \satisf[\ir] \varphi$\quad iff \quad $\modelEpsilon{'},\iota' \satisf[\ir] \varphi$.
\end{enumerate}

{Thus, the reduced models can be used to model-check the \sATLS[\ir] properties of the full models. }

\medskip
\para{Proof idea.}
We aim at showing that the full model $M$ and the reduced one $M'$ satisfy the same formulas of \ATLS[\ir]
referring only to coalitions $\hat{A}\subseteq A$ and containing no nested strategic operators.
Thanks to the restriction on the formulas, the proof can be reduced to showing that $\modelEpsPrim$
satisfies the condition $\AE_A$, which states that for each strategy and for each path of the outcome of this strategy
in $M$ there is an equivalent path in the outcome of the same strategy in $M'$.
In order to show that $\AE_A$ holds, we use the conditions on the selection of events $E(g')$ to be enabled at state $g'$ in $M'$. The conditions include the requirement that $\epsilon$ is always selected, together with the three conditions ${\bf C1, C2, C3}$ adapted from \cite{peled-on_the_fly,cgp99,Jamroga18por}.

Intuitively, ${\bf C1}$ states that, along each path $\pi$ in $M$ which starts at $g'$, each event that is dependent
on an event in $E(g')$ cannot be executed in $M$ unless an event in $E(g')$ is executed first in $M$.
${\bf C2}$ says that $E(g')$ {either contains all the events, or only events that do not change the values of relevant propositions}.
${\bf C3}$ guarantees that for every cycle in $M'$ containing no $\epsilon$-transitions,
there is at least one node $g'$ in the cycle for which all the enabled events of $g'$ are selected.

First, we show that $M$ and $M'$ are stuttering-equivalent, {i.e., they have the same sets of paths modulo stuttering (that is, finite repetition of states on a path)}.
The crucial observation here is that the reduction of $\modelEps$ under the conditions
{\bf C1, C2, C3} is equivalent to the reduction of $\modelEps$ without the $\epsilon$-loops
under the conditions {\bf C1, C2, C3} of~\cite{peled-on_the_fly},
and then adding the $\epsilon$-loops to all the states of the reduced model.
Therefore, for the paths without $\epsilon$-loops the stuttering equivalence
can be shown similarly to~\cite[Theorem~12]{cgp99} while
for the paths with $\epsilon$-loops we need more involved arguments in the proof.
It turns out that in addition to the fact that $M$ and $M'$ are stuttering equivalent,
we can show that stuttering equivalent paths of $M$ and $M'$ have the same maximal
sequence of visible events.
From that, we can prove that $\AE_A$ holds.

%% file: Conclusions.tex
In this paper, we reconsider the asynchronous semantics of strategic ability for multi-agent systems, proposed in~\cite{Jamroga18por}.
We have already hinted at certain problems with the semantics in the extended abstract~\cite{Jamroga21paradoxes-ea}.
Here, we demonstrate in detail how the straightforward combination of strategic reasoning and models of distributed systems leads to counterintuitive interpretation of formulas.
We identify three main sources of problems.
First, the execution semantics does not handle reasoning about deadlock-inducing strategies well.
Secondly, fairness conditions need to be redefined for strategic play.
Thirdly, the class of representations lacks constructions to resolve the tension between the asymmetry imposed by strategic operators on the one hand, and the asymmetry of interaction, e.g., between communicating parties.

{We deal with the problems as follows.
First, we change the execution semantics of strategies in asynchronous MAS by adding ``silent'' $\epsilon$-transitions in states where no ``real'' event can be executed.
We also propose and study the condition of \emph{opponent-reactiveness} that assumes the agents outside the coalition to not obstruct the execution of the strategy forever.
Note that, while the assumption may produce similar interpretation of formulas as in~\cite{Jamroga18por}, it is now explicit -- as opposed to~\cite{Jamroga18por}, where it was ``hardwired'' in the semantics.
The designer or verifier is free to adopt it or reject it, depending on their view of how the agents in the system behave and choose their actions.

Secondly, we propose a new notion of \emph{strategic concurrency-fairness} that selects the fair executions of a strategy.
Thirdly, we allow for nondeterministic choices in agents' repertoires.}
This way, we allow to explicitly specify that one agent has more control over the outcome of an event than the other participants of the event.

The main technical result consists in proving that partial order reduction for strategic abilities~\cite{Jamroga18por} is still correct after the semantic modifications.
Thus, the new, more intuitive semantics admits efficient verification.

\para{Beyond \ATLir.}
In this study, we have concentrated on the logic \ATLsir, i.e., the variant of \ATLs based on memoryless imperfect information strategies.
Clearly, the concerns raised here are not entirely (and not even not primarily) logical.
\ATLsir can be seen as a convenient way to specify the players and the winning conditions in a certain class of games (roughly speaking, $1.5$-player games with imperfect information, positional strategies, and \LTL objectives).
The semantic problems, and our solutions, apply to all such games interpreted over arenas given by asynchronous MAS.

Moreover, most of the claims presented here are not specific to \ir-strategies.
In fact, we conjecture that our examples of semantic side effects carry over to the other types of strategies (except for the existence of coalitions whose all strategies have empty outcomes, which can happen for neither perfect information nor perfect recall).
Similarly, our technical results should carry over to the other strategy types (except for the correctness of POR, which does not hold for agents with perfect information).
We leave the formal analysis of those cases for future work.

\extended{\para{Other issues.}
An interesting question concerns the relationship between asynchronous and synchronous models. We conjecture that AMAS with explicit control can be simulated by concurrent game structures and alternating transition systems. Similarly, it should be possible to simulate CGS and ATS by AMAS with explicit control, at the expense of using a huge space of fully synchronized actions. For the model checking complexity in AMAS with explicit control, we expect the same results as in~\cite{Jamroga20POR-JAIR}.}

%% file: Po-reduction-v3.tex
All the results in this appendix are formulated and proved for the semantics of \ATLir over undeadlocked AMAS with explicit control.
Also, we restrict the formulas to \ATLs without nested strategic modalities and the next step operator $\Next$ (``simple \ATLs'', or \sATLS).
As noted in~\cite{Jamroga18por}, \sATLS is sufficient for most practical specifications and much more expressive than \LTL.
Yet, as we prove below, it enjoys the same efficiency of partial order reduction.

We begin by introducing the relevant notions of equivalence. 
Then, we propose conditions on reduced models that preserve the stuttering equivalence with and without the assumption of \emph{opponent-reactiveness} (\EL). We point out algorithms that generate such models, and prove their correctness.

It should be stressed that the reduction scheme proposed here is general, in the sense that it preserves equivalent representatives of both fair and unfair paths in the model.
In particular, we do \emph{not} propose a variant of POR, optimized for strategic concurrency-fair paths, analogous to reductions of~\cite{peled-representatives} for \CF. A variant of POR for \sATL[\ir] under the \SCF assumption is planned for future work.

\subsection{Properties of Submodels}\label{sec:submodels}

Given an undeadlocked AMAS \epsAMAS, partial order reduction attempts to generate\extended{ only a subset of states and transitions that is sufficient for verification of \epsAMAS, i.e.,} a relevant \emph{submodel} of $\fullmodel(\epsAMAS)$.

\begin{definition}[Submodel]
\label{reducedM}
Let models $\modelEps, \modelEpsPrim$ extend the same AMAS \epsAMAS, so that
$\States' \subseteq \States$,
$\iota \in \States'$,
$T$ is an extension of $T'$,
and $V' = V|_{\States'}$.
Then, we write $\modelEpsPrim \subseteq \modelEps$ and call $\modelEpsPrim$ a \emph{submodel} of $\modelEps$.
\end{definition}
Note that, for each $\state \in \States'$, we have $\Pi_{\modelEpsPrim}(\state) \subseteq \Pi_\modelEps(\state)$.

\begin{lemma}\label{submodel}
Let $\modelEpsPrim \subseteq \modelEps$, $A \in \Agt$, $\strat_A \in \Sigma_A^{\ir}$.
Then, we have
$\outcomeEps_{\modelEpsPrim}(\iota,\strat_A) = \outcomeEps_{\modelEps}(\iota,\strat_A) \cap \Pi_{\modelEpsPrim}(\iota)$.
\end{lemma}
\begin{myproof}
Note that each joint \ir-strategy in $\modelEps$ is also a well defined \ir-joint strategy in $\modelEpsPrim$
as it is defined on the local states of each agent of an AMAS which is extended
by both $\modelEps$ and $\modelEpsPrim$.
The lemma follows directly from the definition of \EL-outcome (Def.~\ref{def:enfliveness} and~\ref{def:outcome3}),
plus the fact that $\Pi_{\modelEpsPrim}(\iota) \subseteq \Pi_{\modelEps}(\iota)$.
\end{myproof}

\begin{lemma}\label{prop:the-same-strategy}
Let $\modelEps$ be a model, $\pi, \pi' \in \Pi_{\modelEps}(\iota)$, and for some $i \in \A:$
$\events(\pi)\mid_{\events_i} = \events(\pi')\mid_{\events_i}$.
Then, for each \ir-strategy $\strat_i$,
we have $\pi \in \outcomeEps_{\modelEps}(\iota,\strat_i)$ iff $\pi' \in \outcomeEps_{\modelEps}(\iota,\strat_i)$.
\end{lemma}
\begin{myproof}
Let $\events(\pi)\mid_{\events_i} = b_0b_1\ldots$ be the sequence of the events of agent $i$ in $\pi$.
For each $b_j$ let $\pi[b_j]$ denote the global state from which $b_j$ is executed in $\pi$.
By induction we can show that for each $j \geq 0$, 
we have $\pi[b_j]^i= \pi'[b_j]^i$.
For $j = 0$ it is easy to see that $\pi[b_0]^i = \pi[b_0]^i = \iota^i$.
Assume that the thesis holds for $j = k$.
The induction step follows from the fact the local evolution $T_i$ is a function,
so if $\pi[b_k]^i = \pi'[b_k]^i = l$ for some $l \in L_i$,
then $\pi[b_{k+1}]^i = \pi'[b_{k+1}]^i = T_i(l,b_k)$.
Thus, by Def.~\ref{def:enfliveness} and ~\ref{def:outcome3}, for each \ir-strategy $\strat_i$
we have $\pi \in \outcomeEps_{\modelEps}(\iota,\strat_i)$
iff $\pi' \in \outcomeEps_{\modelEps}(\iota,\strat_i)$, which concludes the proof.
\end{myproof}

Lemma~\ref{prop:the-same-strategy} can be easily generalized to joint strategies $\strat_A\in\Sigma_A^{\ir}$.

\subsection{Stuttering Equivalence}\label{sec:stuttering-ltlx}

Let $\modelEps$ be a model, $\modelEpsPrim \subseteq \modelEps$, and $\hatPV \subseteq \PV$ a subset of propositions.
Stuttering equivalence says that two paths can be divided into corresponding finite segments,
each satisfying exactly the same propositions.
Stuttering path equivalence\extended{\footnote{
  The property is usually called \emph{stuttering trace equivalence} \cite{cgp99}.
  We use a slightly different name to avoid confusion with Mazurkiewicz traces, also used in this paper. }}
requires two models to always have \extended{corresponding, }stuttering-equivalent paths.

\begin{definition}[Stuttering equivalence]
\label{def-ste}
Two paths $\seq \in \Pi_{\modelEps}(\iota)$ and $\seq' \in \Pi_{\modelEpsPrim}(\iota)$ are {\em stuttering equivalent},
denoted $\seq \equiv_{s} \seq'$,
if there exists a partition $B_0 = (\seq[0],\dots,\seq[i_1-1]),\ B_1=(\seq[i_1],\dots,\seq[i_2-1]),\ \ldots$\ of the states of $\seq$,
and an analogous partition $B'_0, B'_1, \ldots$ of the states of $\seq'$,
s.t. for each $j \geq 0:$ $B_j$ and $B'_j$ are nonempty and finite, and
$V(\state)\cap\hatPV = V'(\state')\cap\hatPV$ for every $\state\in B_j$ and $\state'\in B'_j$.

Models $\modelEps$ and $\modelEpsPrim$ are {\em stuttering path equivalent},
denoted $\modelEps \equiv_{s} \modelEpsPrim$ if for each path $\seq \in \Pi_{\modelEps}(\iota)$,
there is a path $\seq' \in \Pi_{\modelEpsPrim}(\iota)$ such that $\seq \equiv_{s} \seq'$.\footnote{Typically,
the definition also contains the symmetric condition which in our case always holds for $\modelEps$ and its submodel $\modelEpsPrim$,
as $\Pi_{\modelEpsPrim}(\iota) \subseteq \Pi_{\modelEps}(\iota)$. }
\end{definition}

\begin{theorem}[\cite{cgp99}]\label{equivs}
If $\modelEps \equiv_{s} \modelEpsPrim$, then we have $\modelEps, \iota \models \varphi$ iff $\modelEpsPrim, \iota' \models \varphi$,
for any \LTLX\ formula $\varphi$ over $\hatPV$.
\end{theorem}

\subsection{Independence of Events}\label{sec:independence}

Intuitively, an event is invisible iff it does not change the valuations of the propositions.\footnote{
  This concept of invisibility is technical, and is not connected to the view of any agent\extended{ in the sense of~\cite{MalvoneMS17}}. }
Additionally, we can designate a subset of agents $A$ whose events are visible by definition.
Furthermore, two events are independent iff they are not events of the same agent and at least one of them is invisible.

\begin{definition}[Invisible events]
Consider a model $M$, a subset of agents $A\subseteq\A$, and a subset of propositions $\hatPV\subseteq \PV$.
An event $\evt \in \events$ is {\em invisible}
wrt.~$A$ and $\hatPV$ if $Agent(\evt)\cap A = \emptyset$ and for each two global states
$\state, \state' \in \States$ we have that $\state \trans \evt \state'$ implies
$V(\state)\cap\hatPV = V(\state')\cap\hatPV$.
The set of all invisible events for $A,\hatPV$ is denoted by $Invis_{A,\hatPV}$,
and its closure -- of visible events -- by $Vis_{A,\hatPV} = \events \setminus Invis_{A,\hatPV}$.
\end{definition}

\begin{definition}[Independent events]
The notion of \emph{independence} $I_{A,\hatPV}\subseteq \events\times \events$ is defined as:
$I_{A,\hatPV} = \{(\evt,\evt') \in \events \times \events \mid Agent(\evt) \cap Agent(\evt') = \emptyset\}\ \setminus\ (Vis_{A,\hatPV} \times Vis_{A,\hatPV})$.
Events $\evt, \evt' \in \events$ are called {\em dependent} if $(\evt,\evt') \not \in I_{A,\hatPV}$.
If it is clear from the context, we omit the subscript $\hatPV$.
\end{definition}

\subsection{Preserving Stuttering Equivalence} 
\label{sec:POR-correctness}

Rather than generating the full model $\modelEps=\fullmodel(\epsAMAS)$, one can generate a reduced model $\modelEpsPrim$
satisfying the following property:

\begin{center}
$\AE_A:\:
 \forall \strat_A\!\in\!\Sigma_A^\ir\quad
 \forall \seq\! \in\! \outcomeEps_{\modelEps}(\iota,\strat_A)$ \\
$\qquad \exists \seq'\! \in\! \outcomeEps_{\modelEpsPrim}(\iota,\strat_A)\quad
    \seq\! \equiv_s\! \seq'$.
\end{center}

We define a class of algorithms that generate reduced models satisfying $\AE_A$ (Section~\ref{sec:PORA}),
and then prove that these models preserve \sATLS[\ir] (Section~\ref{sec:POR-correctness-AEA}).

\para{Algorithms for partial order reduction.}\label{sec:PORA}
POR is used to reduce the size of models while preserving satisfaction for a class of formulas.
The standard DFS~\cite{GKPP99} or DDFS~\cite{CVWY92} is modified in such a way that from each visited state $g$
an event $\evt$ to compute the successor state $g_1$ such that $g \stackrel{\evt}{\rightarrow} g_1$,
is selected from $E(g)\cup\set{\epsilon}$ such that $E(g) \subseteq enabled(g)\setminus\set{\epsilon}$. That is, the algorithm always selects $\epsilon$, plus a subset of the enabled events at $g$.
Let $A \subseteq \A$.
The conditions on the heuristic selection of $E(g)$ given below are inspired by~\cite{peled-on_the_fly,cgp99,Jamroga18por}.
\begin{description}
\item[{\bf C1}]
    Along each path $\pi$ in $\modelEps$ that starts at $g$, each event that is dependent 
	on an event in $E(g)$ cannot be executed in $\pi$ without an event in $E(g)$ being executed first in $\pi$.
	Formally, $\forall \pi \in \Pi_{\modelEps}(g)$ such that $\pi = g_0\evt_0g_1\evt_1\ldots$ with $g_0 = g$,
	and $\forall b \in \events$ such that $(b,c) \notin I_A$ for some $c \in E(g)$,
	if $\evt_i = b$ for some $i \geq 0$, then $\evt_j \in E(g)$ for some $j < i$.
\item[{\bf C2}]
    If $E(g) \neq enabled(g)\setminus\set{\epsilon}$, then $E(g) \subseteq Invis_A$. 
\item[{\bf C3}]
    For every cycle in $\modelEpsPrim$ containing no $\epsilon$-transitions, there is at least one node $g$ in the cycle for which $E(g) = enabled(g)\setminus\set{\epsilon}$,
		i.e., for which all the successors of $g$ are expanded.
\end{description}

\begin{theorem}\label{prop:stequ}
Let $A \subseteq \A$, $\modelEps=\fullmodel(\epsAMAS)$, and $\modelEpsPrim \subseteq \modelEps$ be the reduced model generated by
DFS with the choice of $E(g')$ for $g' \in \States'$ given by conditions {\bf C1, C2, C3}
and the independence relation $I_A$. Then, $\modelEpsPrim$ satisfies $\AE_A$.
\end{theorem}
\begin{myproof}
Let $\modelEpsPrim \subseteq \modelEps=\fullmodel(\epsAMAS)$ be the reduced model generated as specified.
Notice that the reduction of $\modelEps$ under the conditions {\bf C1, C2, C3} above is equivalent
to the reduction of $\modelEps$ without the $\epsilon$-loops under the conditions
{\bf C1, C2, C3} of~\cite{peled-on_the_fly},
and then adding the $\epsilon$-loops to all the states of the reduced model.
Although the setting is slightly different, it can be shown similarly to~\cite[Theorem~12]{cgp99}
that the conditions {\bf C1, C2, C3} guarantee that the models:
(i) $\modelEps$ without $\epsilon$-loops and (ii) $\modelEpsPrim$ without $\epsilon$-loops are stuttering path equivalent.
More precisely, for each path $\pi = g_0a_0g_1a_1\cdots$ with $g_0 = \iota$ (without $\epsilon$-transitions) in $M$
there is a stuttering equivalent path $\pi' = g'_0a'_0g'_1a'_1\cdots$ with $g'_0 = \iota$ (without $\epsilon$-transitions) in $M'$ such that
$\events(\pi)|_{Vis_A} = \events(\pi')|_{Vis_A}$,
i.e., $\pi$ and $\pi'$ have the same maximal sequence of visible events for $A$. \textbf{(*)}

We will now prove that this implies $\modelEps \equiv_s \modelEpsPrim$.
Removing the $\epsilon$-loops from $\modelEps$ eliminates two kinds of paths:
(a) paths with infinitely many ``proper'' events, and
(b) paths ending with an infinite sequence of $\epsilon$-transitions.
Consider a path $\seq$ of type (a) from $\modelEps$.
Notice that the path $\seq_1$, obtained by removing the $\epsilon$-transitions from $\seq$,
is stuttering-equivalent to $\seq$.
Moreover,  by~\textbf{(*)}, there exists a path $\seq_2$ in $\modelEpsPrim$ without $\epsilon$-transitions,
which is stuttering-equivalent to $\seq_1$.
By transitivity of the stuttering equivalence, we have that $\seq_2$ is stuttering equivalent to $\seq$.
Since $\seq_2$ must also be a path in $\modelEpsPrim$, this concludes this part of the proof.

Consider a path $\seq$ of type (b) from $\modelEps$, i.e., $\seq$ ends with an infinite sequence of $\epsilon$-transitions.
Let $\seq_1$ be the sequence obtained from $\seq$ after removing $\epsilon$-transitions,
and $\seq_2$ be any infinite path without $\epsilon$-transitions such that $\seq_1$ is its prefix.
Then, it follows from~\textbf{(*)} that there is a stuttering equivalent path
$\seq_2' = g'_0a'_0g'_1a'_1\cdots$ with $g'_0 = \iota$ in $M'$ such that $\events(\seq_2)|_{Vis_A} = \events(\seq_2')|_{Vis_A}$.
Consider the minimal finite prefix $\seq_1'$ of $\seq_2'$ such that  $\events(\seq_1')|_{Vis_A} = \events(\seq_1)|_{Vis_A}$.
Clearly, $\seq_1'$ is a sequence in $M'$ and can be extended with an infinite number of $\epsilon$-transitions to the path $\seq'$ in $M'$.
It is easy to see that $\seq$ and $\seq'$ are stuttering equivalent.

So far, we have shown that our reduction under the conditions {\bf C1, C2, C3} guarantees that the models
$\modelEps$ and $\modelEpsPrim$ are stuttering path equivalent, and more precisely
that for each path $\pi = g_0a_0g_1a_1\cdots$ with $g_0 = \iota$ in $M$
there is a stuttering equivalent path $\pi' = g'_0a'_0g'_1a'_1\cdots$ with $g'_0 = \iota$ in $M'$ such that
$\events(\pi)|_{Vis_A} = \events(\pi')|_{Vis_A}$,
i.e., $\pi$ and $\pi'$ have the same maximal sequence of visible events for $A$.
To show that $\modelEpsPrim$ satisfies $\AE_A$, consider an \ir-joint strategy $\strat_A$ and $\seq \in \outcomeEps_\modelEps(\iota,\strat_A)$.
As demonstrated above, there is $\seq' \in \Pi_{\modelEpsPrim}(\iota)$ such that $\seq \equiv_s \seq'$
and $\events(\seq)|_{Vis_A} = \events(\seq')|_{Vis_A}$.
Since $\events_i \subseteq Vis_A$ for each $i \in A$, the same sequence of events of each $\events_i$ is
executed in $\seq$ and $\seq'$.
Thus, by the generalization of Lemma~\ref{prop:the-same-strategy} to \ir-joint strategies we get
$\seq' \in \outcomeEps_\modelEps(\iota,\strat_A)$.
So, by Lemma~\ref{submodel} we have $\seq' \in \outcomeEps_\modelEpsPrim(\iota,\strat_A)$.
\end{myproof}

Algorithms generating reduced models, in which the choice of $E(g)$ is given by similar conditions, can be found for instance in~\short{\cite{peled-on_the_fly,peled-representatives,cgp99,GKPP99,LomuscioPQ10b}}%
\extended{\cite{peled-on_the_fly,peled-representatives,cgp99,GKPP99,PenczekSGK00,LomuscioPQ10b}}.

\para{POR for proactive opponents.}\label{sec:PORA-NoReact}
\input{sketch}

\para{Correctness of reductions satisfying $\AE_A$.}\label{sec:POR-correctness-AEA}
We show that the reduced models satisfying $\AE_A$ preserve \sATLS[\ir].

\begin{theorem}\label{prop:ae-atlir}
Let $A\subseteq\A$, and let models $\modelEpsPrim \subseteq \modelEps$, $\modelEpsilon{'} \subseteq \modelEpsilon$ satisfy $\AE_A$.
For each \sATLS[\ir] formula $\varphi${ over $\hatPV$,} that refers only to coalitions $\hat{A}\subseteq A$,
we have that:
\begin{enumerate}
\item $\modelEps,\iota \satisfair[\ir]{\EL} \varphi$\quad iff \quad $\modelEpsPrim,\iota' \satisfair[\ir]{\EL} \varphi$,\quad and
\item $\modelEpsilon,\iota \satisf[\ir] \varphi$\quad iff \quad $\modelEpsilon{'},\iota' \satisf[\ir] \varphi$.
\end{enumerate}
\end{theorem}
\begin{myproof}
Proof by induction on the structure of $\varphi$.
We show the case $\varphi = \coop{\hat{A}}\gamma$.
The cases for $\neg, \land$ are straightforward.

Notice that $\outcomeEps_{\modelEpsPrim}(\iota,\strat_{\hat{A}}) \subseteq \outcomeEps_{\modelEps}(\iota,\strat_{\hat{A}})$,
which together with \extended{the condition }$\AE_A$ implies that the sets $\outcomeEps_{\modelEps}(\iota,\strat_{\hat{A}})$ and $\outcomeEps_{\modelEpsPrim}(\iota,\strat_{\hat{A}})$ are stuttering path equivalent. Analogously, this is the case for $\outcome_{\modelEpsPrim}(\iota,\strat_{\hat{A}}) \subseteq \outcome_{\modelEps}(\iota,\strat_{\hat{A}})$, i.e. without the \EL assumption.
Hence, (1) and (2) follow from Theorem \ref{equivs}.
\end{myproof}

Together with Theorems~\ref{prop:stequ} and~\ref{prop:withoutEL}, we obtain the following.
\begin{theorem}\label{prop:ae-atlir-corollary}
Let $\modelEps=\fullmodel(\epsAMAS)$, $\modelEpsilon=\IISEps(\AMAS)$ and let $\modelEpsPrim \subseteq \modelEps$ and $\modelEpsilon{'} \subseteq \modelEpsilon$ be the reduced models generated by
DFS with the choice of $E(g')$ for $g' \in \States'$ given by conditions {\bf C1, C2, C3}
and the independence relation $I_{A,\hatPV}$.
For each \sATLS[\ir] formula $\varphi$ over $\hatPV$, that refers only to coalitions $\hat{A}\subseteq A$,
we have:
\begin{enumerate}
\item $\modelEps,\iota \satisfair[\ir]{\EL} \varphi$\quad iff \quad $\modelEpsPrim,\iota' \satisfair[\ir]{\EL} \varphi$,\quad and
\item $\modelEpsilon,\iota \satisf[\ir] \varphi$\quad iff \quad $\modelEpsilon{'},\iota' \satisf[\ir] \varphi$.
\end{enumerate}
\end{theorem}

This concludes the proof that the adaptation of POR for \LTLX to \sATLS[\ir], originally presented in~\cite{Jamroga18por},
remains sound in the updated semantics proposed in Sections~\ref{sec:paradoxes} and \ref{sec:asymmetric}.
That is, the structural condition $\AE_A$ is sufficient to obtain correct reductions for \sATLS[\ir] with and without the new opponent-reactiveness assumption (Theorem~\ref{prop:ae-atlir-corollary}).
Thanks to that, one can potentially reuse or adapt the existing POR algorithms and tools for \LTLX, and the actual reductions are likely to be substantial.

%% file: Sketch.tex
The same reduction still works without the assumption of opponent-reactiveness (\EL).

\begin{theorem}\label{prop:withoutEL}
Let $\modelEpsilon = \IISEps(\AMAS)$ be an undeadlocked IIS.
Then, its reduced model $\modelEpsilon{'}$, generated as in Theorem~\ref{prop:stequ}, satisfies $\AE_A$.
\end{theorem}
\begin{myproof}[ (Sketch)] 
In this setting, there is no auxiliary agent in the AMAS, and $\epsilon$-transitions are added directly to the IIS in accordance with Definition~\ref{def:undeadlockedIIS}.
Hence, not every global state of $\modelEpsilon$ necessarily has an $\epsilon$ loop, but only those where a miscoordinating combination of events exists.
However, this does not impact the reduction itself.

First, note that Lemma~\ref{submodel} still holds, directly from the definition of outcome (Definition~\ref{def:outcome}).
Furthermore, because in the undeadlocked IIS $\modelEpsilon$ the $\epsilon$-transitions do not belong to any agent,
Lemma~\ref{prop:the-same-strategy}, where sequences of some agent $i$'s events are considered, also holds.
Note that the \EL condition only restricts the outcome sets, and not the model itself:
both $\modelEps = IIS(\AMAS^\epsilon)$ and $\modelEpsilon$ contain the same two types (a) and (b) of paths with $\epsilon$-transitions as discussed in Theorem~\ref{prop:stequ}.
Hence, following its reasoning, it can first be shown that models $\modelEpsilon$ and $\modelEpsilon{'}$ without their $\epsilon$-transitions are stuttering path equivalent,
and that it remains the case also when both types of paths including $\epsilon$ loops are included.

Note that the remark about $\modelEpsPrim$ being equivalent to reducing $\modelEps$ without $\epsilon$ loops and adding them to each global state obviously does not apply to $\modelEpsilon$
(not every global state of $\modelEpsilon$ has them in the first place).
However, this observation has no bearing on the proof.
As before, $\epsilon$ is explicitly stated to be selected for the subset $E(g)$,
ensuring preservation of representative paths with $\epsilon$ in $\modelEpsilon{'}$.
\end{myproof}